\documentclass[10pt,a4paper]{article}
\usepackage[top=1.5cm, bottom=1.5cm, left=1.5cm, right=1.5cm]{geometry} 
\usepackage{graphicx}
\usepackage{amsmath, amssymb, amsfonts}
\usepackage{nicefrac}
\usepackage{color}
\usepackage{array}
\usepackage{bm}
\usepackage{multirow}
\usepackage{abstract}
\usepackage{authblk}  
\usepackage{comment}
\usepackage[utf8]{inputenc}    
\usepackage[T1]{fontenc}
\renewcommand{\arraystretch}{1.2}

\begin{document}
\title{Annealing approximation in master-node network model}
\author[1]{M. O. Hase}
\author[2]{Anderson A. Ferreira}
\author[1]{André C. R. Martins}
\author[1,3]{Fernando F. Ferreira}

\affil[1]{School of Arts, Sciences, and Humanities, University of S\~ao Paulo, S\~ao Paulo, SP 03828-000, Brazil}
\affil[2]{Department of Physics, Federal University of S\~ao Paulo, Diadema, SP 09972-270, Brazil}
\affil[3]{Department of Physics, FFCLRP, University of S\~ao Paulo, S\~ao Paulo, SP 14040-901, Brazil}

\date{\today}

\maketitle

\begin{abstract}
This paper investigates absorbing-state phase transitions in opinion dynamics through a master-node network model analyzed using annealing approximation. We develop a theoretical framework examining three fundamental regimes: systems converging to complete disagreement, complete consensus, or both states depending on initial conditions. The phase behavior is governed by two key chiral parameters: $R$ measuring right-oriented influence and $L$ measuring left-oriented influence in the network interactions. Our analysis reveals a rich phase diagram featuring both continuous and discontinuous transitions between disordered and ordered phases. The discontinuous transition emerges in systems with two absorbing states, where the final configuration depends critically on initial opinion distributions. The annealing approximation provides fundamental insights into how asymmetric social influences (chirality) shape collective opinion formation, acting as a symmetry-breaking element that drives the system toward polarization or consensus.
\end{abstract}

\section{Introduction}

The study of absorbing-state transitions provides fundamental insights into opinion dynamics within social systems. These transitions occur when a system evolves into a stable configuration from which it cannot escape, representing either consensus formation or the dominance of a particular opinion among agents. This phenomenon is particularly relevant for understanding how opinions emerge, propagate, and stabilize in populations, revealing the mechanisms through which individuals influence each other and converge toward shared beliefs. In social contexts, absorbing states represent conditions where opinions become fixed, with agents unable to return to previous belief states. This process is crucial for understanding consensus formation, where groups reach stable agreements shaped by social interactions, cultural influences, and external pressures \cite{galam1991towards,galam2022opinion,fridman2013impact,van2024social}.

Models of opinion formation demonstrate how individuals modify their beliefs through peer interactions, leading to either consensus or polarization \cite{bhat2020polarization,battiston2016interplay}. A critical aspect of these dynamics involves the interplay between individual cognitive processes and broader sociocultural contexts, which provide frameworks for interpreting and responding to social cues \cite{battiston2016interplay,galesic2021integrating}. Absorbing-state phase transitions, where systems become trapped in particular configurations, offer valuable perspectives on how individuals mutually influence each other and converge toward collective beliefs \cite{castellano2009statistical,pires2022double}. The critical properties observed in these transitions suggest universal behavior across diverse network structures, with local interactions playing a pivotal role in determining outcomes \cite{hinrichsen2000non}.

Cultural factors significantly shape these dynamics by embedding norms and values into interaction processes, thereby influencing the trajectory of opinion evolution \cite{morris2015normology,crokidakis2024nonequilibrium}. Understanding these mechanisms is essential for explaining how stable opinion patterns emerge in societies, revealing the complex relationships between individual beliefs, social influence, and cultural context \cite{moussaid2013social,javarone2014social}. Such insights extend beyond technical modeling, providing a comprehensive framework for analyzing how social systems stabilize and evolve.

Recent developments in opinion dynamics have introduced master-node models to investigate multi-agent systems influenced by central nodes within networked structures \cite{ferreira2020stochastic,mihara2023critical}. The work of \cite{ferreira2020stochastic} examines how a master node with preferential opinions affects consensus formation and social dynamics. Using stochastic quenched disorder models and mean-field approximations, the study reveals a rich phase diagram connecting critical parameters with collective behavior transitions, offering insights into mechanisms underlying morality, innovation, and cooperation in social systems. Complementing this, \cite{mihara2023critical} conducts rigorous analysis of critical exponents in one-dimensional networks through Monte Carlo simulations and finite-size scaling methods. The research identifies a continuous phase transition to an absorbing state that does not conform to known universality classes, highlighting the model's unique properties.

These studies provide complementary perspectives on networked agent dynamics under master node influence. While \cite{ferreira2020stochastic} establishes the foundational framework and demonstrates phase transitions as key mechanisms in collective behavior, \cite{mihara2023critical} offers detailed characterization of critical properties through exponent analysis. Together, they advance understanding of decentralized decision-making processes while contributing to nonequilibrium statistical mechanics. The current work builds upon these foundations by introducing a chiral master node formulation that incorporates geometric properties and Berry phase effects, enabling more comprehensive analysis of symmetry breaking patterns in opinion dynamics. The general formulation of the model, as shown in Section 2, contains so many parameters that it leads to a system that is not analytically accessible at first sight. Nevertheless, introducing an annealing approximation (on the variable associated with the influence of the master node) and examining the model at the mean-field level generates a system where several relevant cases can be characterized by just two quantities, as shown in Section 3. The analysis is then organized into three parts (Sections 4 to 6) and the role of the master node in this approximation is discussed in Section 7. Some final remarks can be found in Section 8. 

\section{General formulation (continuous approach)}

\begin{figure}[ht]
\centering
\includegraphics[width=100pt]{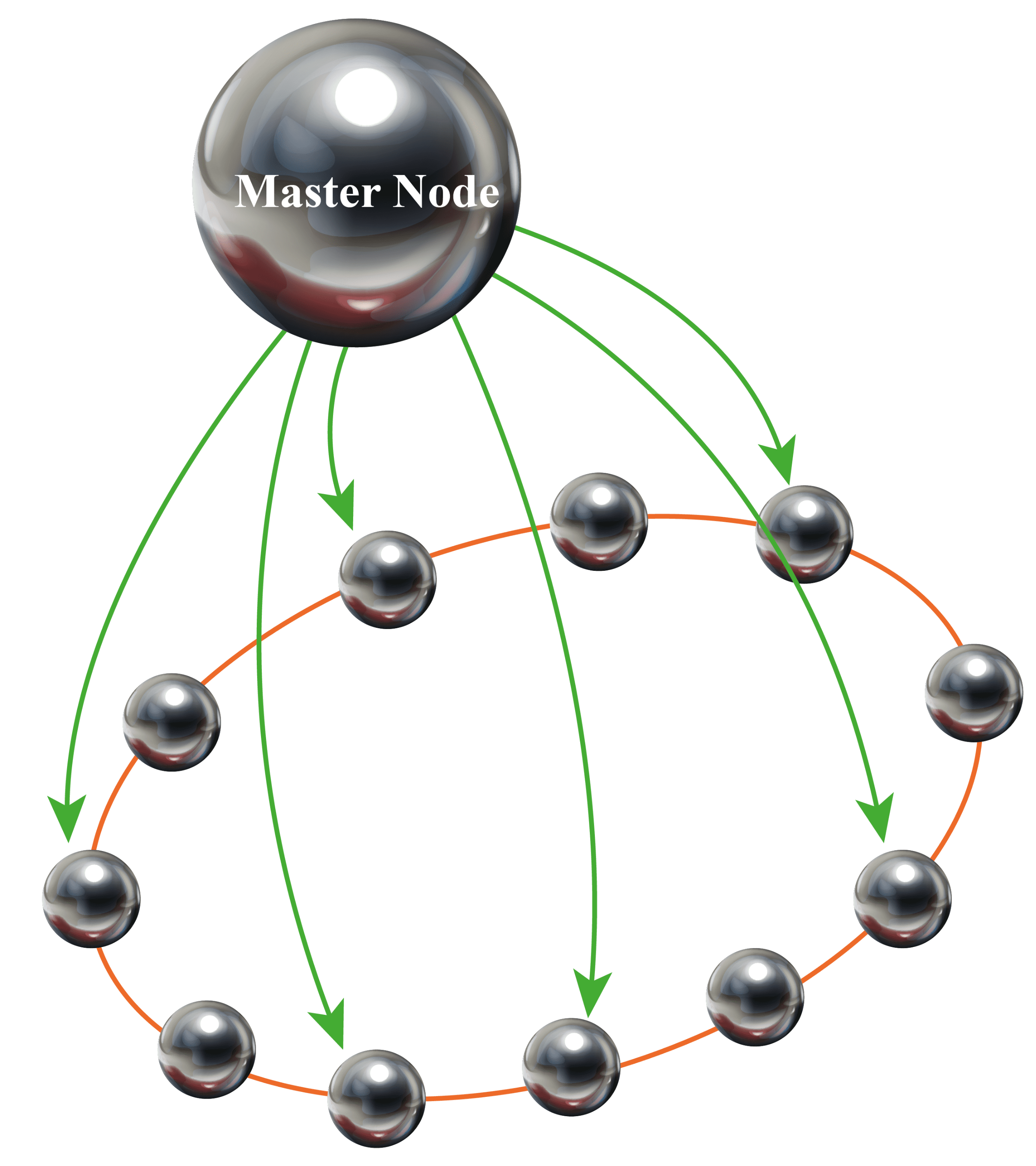}  
\caption{Sketch of the system.}
\label{ringmasternode}
\end{figure}

The model consists of a one-dimensional chain where an individual (or spin) is assigned to each one of its $N$ sites. We denote by $s_{i}$ the state (or opinion) of individual $i$, who is located at position $i\in\{1,\ldots,N\}$ by construction. The periodic boundary condition is adopted ($s_{N+1}=s_{1}$) and the system can be seen as a \textquotedblleft ring\textquotedblright, as shown in Figure \ref{ringmasternode}. The spins have short-ranged interaction in the sense that an individual $i$ interacts with spins $i-1$ and $i+1$ only. A master node is then introduced to this setup and it can influence the way an individual interacts with its neighbors. However the master node is not affected by the state of the $N$ individuals of the ring.

The state of the above-defined model is completely described by the vector $s:=(s_{1}, \ldots, s_{N})$, where $s_{i}\in\{0,1\}$ for each one of the $N$ vertices of the model. The dynamics is governed by the master equation
\begin{align}
\frac{\textup{d}}{\textup{d}t}P(s, t) = \sum_{i=1}^{N}\left[ w_{i}(s^{i})P(s^{i}, t) - w_{i}(s)P(s, t) \right],
\label{me_cont}
\end{align}
which characterizes the time evolution of the probability $P$ of the system begin at state $s$ at time $t$. In \eqref{me_cont}, $s^{i}:=(s_{1}, \ldots, 1-s_{i}, \cdots, s_{N})$ denotes the state where the $i$th state is replaced by its complement and $w_{i}$ is the transition rate (there should be no confusion with the parameters $w_{1,2,3,4}$ introduced in the model) of the individual at site $i$.


\begin{table}[h!]
\centering
\begin{tabular}{|c|c|c|c|c|c|}\hline
\multicolumn{2}{|c}{Neighbors} & \multicolumn{4}{|c|}{$s_{i}\rightarrow s_{i}^{\prime}$ \quad ($\Gamma_{i}\in\{0,1\}$)} \\\hline\hline
$s_{i-1}$ & $s_{i+1}$ & $0\rightarrow 0$ & $0\rightarrow 1$ & $1\rightarrow 0$ & $1\rightarrow 1$ \\\hline 
$0$ & $0$ & $p_{1}+\Gamma_{i}\left(q_{1}-p_{1}\right)$ & $\overline{p_{1}}+\Gamma_{i}\left(\overline{q_{1}}-\overline{p_{1}}\right)$ & $p_{2}+\Gamma_{i}\left(q_{2}-p_{2}\right)$ & $\overline{p_{2}}+\Gamma_{i}\left(\overline{q_{2}}-\overline{p_{2}}\right)$ \\\hline 
$1$ & $1$ & $p_{3}+\Gamma_{i}\left(q_{3}-p_{3}\right)$ & $\overline{p_{3}}+\Gamma_{i}\left(\overline{q_{3}}-\overline{p_{3}}\right)$ & $p_{4}+\Gamma_{i}\left(q_{4}-p_{4}\right)$ & $\overline{p_{4}}+\Gamma_{i}\left(\overline{q_{4}}-\overline{p_{4}}\right)$ \\\hline 
$0$ & $1$ & $w_{1}+\Gamma_{i}\left(v_{1}-w_{1}\right)$ & $\overline{w_{1}}+\Gamma_{i}\left(\overline{v_{1}}-\overline{w_{1}}\right)$ & $w_{2}+\Gamma_{i}\left(v_{2}-w_{2}\right)$ & $\overline{w_{2}}+\Gamma_{i}\left(\overline{v_{2}}-\overline{w_{2}}\right)$ \\\hline 
$1$ & $0$ & $w_{3}+\Gamma_{i}\left(v_{3}-w_{3}\right)$ & $\overline{w_{3}}+\Gamma_{i}\left(\overline{v_{3}}-\overline{w_{3}}\right)$ & $w_{4}+\Gamma_{i}\left(v_{4}-w_{4}\right)$ & $\overline{w_{4}}+\Gamma_{i}\left(\overline{v_{4}}-\overline{w_{4}}\right)$ \\\hline
\end{tabular}
\caption{Transition rates for all possible neighbor configurations $(s_{i-1}, s_{i+1})$ and state changes $(s_i\rightarrow s_i')$. The rates depend on both the base parameters ($p_j$, $w_j$, $v_j$, $q_j$) and the master node influence $\Gamma_i$. The overline notation $\overline{x}$ denotes the complementary probability $1-x$.}
\label{t1}
\end{table}

 Following the rules presented in table \ref{t1}, the transition rate can be represented by
\begin{align}
\nonumber w_{i}(s) &= s_{i}\big\{ \left(1-s_{i-1}\right)\left(1-s_{i+1}\right)\left[p_{2}+\Gamma_{i}\left(q_{2}-p_{2}\right)\right] + s_{i-1}s_{i+1}\left[p_{4}+\Gamma_{i}\left(q_{4}-p_{4}\right)\right] \\
\nonumber &+ \left(1-s_{i-1}\right)s_{i+1}\left[w_{2}+\Gamma_{i}\left(v_{2}-w_{2}\right)\right] + s_{i-1}\left(1-s_{i+1}\right)\left[w_{4}+\Gamma_{i}\left(v_{4}-w_{4}\right)\right] \big\} \\
\nonumber &+ \left(1-s_{i}\right)\big\{ \left(1-s_{i-1}\right)\left(1-s_{i+1}\right)\left[\overline{p_{1}}+\Gamma_{i}\left(\overline{q_{1}}-\overline{p_{1}}\right)\right] + s_{i-1}s_{i+1}\left[\overline{p_{3}}+\Gamma_{i}\left(\overline{q_{3}}-\overline{p_{3}}\right)\right] \\
\nonumber &+ \left(1-s_{i-1}\right)s_{i+1}\left[\overline{w_{1}}+\Gamma_{i}\left(\overline{v_{1}}-\overline{w_{1}}\right)\right] + s_{i-1}\left(1-s_{i+1}\right)\left[\overline{w_{3}}+\Gamma_{i}\left(\overline{v_{3}}-\overline{w_{3}}\right)\right] \big\} \\
\nonumber &= s_{i}\left[A_{i}^{(2)} + \left(B_{i}^{(4)}-A_{i}^{(2)}\right)s_{i-1} + \left(B_{i}^{(2)}-A_{i}^{(2)}\right)s_{i+1} + \left(A_{i}^{(2)}+A_{i}^{(4)}-B_{i}^{(2)}-B_{i}^{(4)}\right)s_{i-1}s_{i+1}\right] \\
 &+ \left(1-s_{i}\right)\left[\overline{A_{i}^{(1)}} + \left(\overline{B_{i}^{(3)}}-\overline{A_{i}^{(1)}}\right)s_{i-1} + \left(\overline{B_{i}^{(1)}}-\overline{A_{i}^{(1)}}\right)s_{i+1} + \left(\overline{A_{i}^{(1)}}+\overline{A_{i}^{(3)}}-\overline{B_{i}^{(1)}}-\overline{B_{i}^{(3)}}\right)s_{i-1}s_{i+1}\right],
\label{wi}
\end{align}
where
\begin{align}
A_{i}^{(j)} := p_{j} + \Gamma_{i}\left(q_{j} - p_{j}\right),\quad \overline{A_{i}^{(j)}} := \overline{p_{j}} + \Gamma_{i}\left(\overline{q_{j}}-\overline{p_{j}}\right),\quad B_{i}^{(j)} := w_{j} + \Gamma_{i}\left(v_{j} - w_{j}\right)\;\text{ and }\;\overline{B_{i}^{(j)}} := \overline{w_{j}} + \Gamma_{i}\left(\overline{v_{j}}-\overline{w_{j}}\right)
\label{AB}
\end{align}
for $j\in\{1,2,3,4\}$.

The magnetization of the $i$th individual is defined as usual,
\begin{align}
m_{i}(t) := \sum_{s}s_{i}P(s,t),
\label{m}
\end{align}
where the sum runs over the set $\{0,1\}^{N}$. From the master equation \eqref{me_cont}, the dynamics of the magnetization is described by
\begin{align}
\frac{\textup{d}}{\textup{d}t}m_{i}(t) := \sum_{s}\left(1-2s_{i}\right)w_{i}(s)P(s,t),
\label{dm/dt}
\end{align}
and from the particular form \eqref{wi}, it implies
\begin{align}
\nonumber\frac{\textup{d}}{\textup{d}t}m_{i}(t) &= \overline{A_{i}^{(1)}} - \left(\overline{A_{i}^{(1)}}+A_{i}^{(2)}\right)m_{i}(t) + \left(\overline{B_{i}^{(3)}}-\overline{A_{i}^{(1)}}\right)m_{i-1}(t) + \left(\overline{B_{i}^{(1)}}-\overline{A_{i}^{(1)}}\right)m_{i+1}(t) + \\
\nonumber &+ \left(\overline{A_{i}^{(1)}}+A_{i}^{(2)}-\overline{B_{i}^{(3)}}-B_{i}^{(4)}\right)q_{i,i-1}(t) + \left(\overline{A_{i}^{(1)}}+A_{i}^{(2)}-\overline{B_{i}^{(1)}}-B_{i}^{(2)}\right)q_{i,i+1}(t) + \\
\nonumber &+ \left(\overline{A_{i}^{(1)}}+\overline{A_{i}^{(3)}} - \overline{B_{i}^{(1)}} - \overline{B_{i}^{(3)}}\right)q_{i-1,i+1}(t) + \\
 &+ \left(\overline{B_{i}^{(1)}} + B_{i}^{(2)} + \overline{B_{i}^{(3)}} + B_{i}^{(4)} - \overline{A_{i}^{(1)}} - A_{i}^{(2)} - \overline{A_{i}^{(3)}} - A_{i}^{(4)}\right)c_{i-1,i,i+1}(t),
\label{m_cont}
\end{align}
where
\begin{align}
q_{i,j}(t) := \sum_{s}s_{i}s_{j}P(s, t) \;\textnormal{ and }\; c_{i,j,k}(t) := \sum_{s}s_{i}s_{j}s_{k}P(s, t).
\label{qc}
\end{align}


\section{Annealed mean-field analysis}

The usual assumptions
\begin{align}
m_{i}(t) = m(t),\quad q_{i,j}(t) = m^{2}(t)\quad (i\neq j),\quad\text{ and }\quad c_{i,j,k}(t) = m^{3}(t)\quad (i\neq j\neq k\neq i),
\label{simple_mf}
\end{align}
are introduced to investigate the dynamics \eqref{m_cont} at the mean-field level. Furthermore, the annealed approach is assumed for the random variable $\Gamma_{i}$, which is then replaced by its expected value $\Gamma_{i}\rightarrow\langle\Gamma_{i}\rangle=\rho$, which is the fraction of connections from the master node. In this annealed approximation, the timescale of the fluctuation of the random variable $\Gamma_{i}$ is assumed to be much faster than the other variables of the system, which means that the influence of the master node over the individuals is felt by an effective one. From a social perspective, we can illustrate this scenario with the following example. Suppose that a central agent (such as mass media) controls the way the population interacts, and let the strength of this influence be represented by a real number in $[0,1]\subset\mathbb{R}$. In the so-called quenched case, the central agent influences only a fraction $\rho$ of the population, but each affected individual experiences the full impact, with strength $1$. In the annealed scenario, the whole population is influenced, but the strength of the effect is just $\rho\in[0,1]$ for each individual.

As a consequence of these hypothesis (annealed and mean-field approach), all the subindices of $A_{i}^{(j)}\rightarrow A^{(j)}$, $\overline{A_{i}^{(j)}}\rightarrow \overline{A^{(j)}}$, $B_{i}^{(j)}\rightarrow B^{(j)}$ and $\overline{B_{i}^{(j)}}\rightarrow \overline{B^{(j)}}$ are now irrelevant and are going to be omitted. Furthermore, the dependence of $m(t)$ on time is also suppressed from now on in the notation.

Taking into account these approximations, it is straightforward to show that the dynamics can now be described by
\begin{align}
\dot m = F(m),
\label{dmdtF}
\end{align}
where $\dot m$ stands for the time derivative of $m$ and
\begin{align}
F(m) := \overline{A^{(1)}} - \left(3\overline{A^{(1)}} + L\right)m + \left(3\overline{A^{(1)}} + R + 2L\right)m^{2} - \left(R + L + \overline{A^{(1)}} + A^{(4)}\right)m^{3}.
\label{F_cont}
\end{align}
The term \eqref{F_cont} can be regarded as the free-energy in dynamical Landau theory. Here, the quantities $L$ and $R$ are defined as
\begin{align}
L := A^{(2)} - \overline{B^{(1)}} - \overline{B^{(3)}} \quad\text{ and }\quad R:= \overline{A^{(3)}} - B^{(2)} - B^{(4)}
\label{LR}
\end{align}
and play a major role in the analysis of the system and can be interpreted in the following way. The quantities $R$ and $L$ emerge as fundamental parameters representing competing directional influences in the network. However, they do not correspond solely to movements towards one choice or another, as there are positive and negative terms in each of them. And yet, physically, a larger $R$ always corresponds to a right-oriented influence, combining the tendency to adopt right-neighbor opinions against resistance to such adoption. Conversely, $L$ quantifies left-oriented influence through similar mechanisms. Their ratio determines the system's chiral character, with $R>L$ favoring right-biased consensus and vice versa. These parameters capture the inherent asymmetry in opinion propagation through the network. Interestingly, depending on those quantities, we observe a fixed point with coexistence of both opinions, like previous models with inflexibles \cite{galam2007role,martins2013building} or with asymmetric influences \cite{galam2004dynamics}.

Note that, within this approach, the original model, which depends on more than 10 free parameters, can be cast as a function of only four ($R$, $L$, $\overline{A^{(1)}}$ and $A^{(4)}$), thereby rendering it amenable to analytical investigation. At this point, it is interesting to restrict the analysis where some well-known absorbing states are present in the system, namely the stationary states $m^{\ast}=0$ and/or $m^{\ast}=1$ in the solution of \eqref{dmdtF}. From \eqref{F_cont}, one can see that they correspond to the assumption $\overline{A^{(1)}}=0$ and $A^{(4)}=0$, respectively. In what follows, the analysis is divided into the cases shown in Table \ref{tablecases123}.

\begin{table}[h]
\renewcommand*{\arraystretch}{1.4}
\centering
\begin{tabular}{|c|c|}\hline
Case I & $m^{\ast}=0$ is an absorbing state but $m^{\ast}=1$ is not \\\hline
Case II & $m^{\ast}=1$ is an absorbing state but $m^{\ast}=0$ is not \\\hline
Case III & $m^{\ast}=0$ and $m^{\ast}=1$ are both absorbing states \\\hline
\end{tabular}
\caption{Cases based on the absorbing states.}
\label{tablecases123}
\end{table}


\section{System with absorbing state $m^{\ast}=0$ (without the absorbing state $m^{\ast}=1$): Case I}

\begin{figure}[ht!]
\centering
\includegraphics[width=346pt]{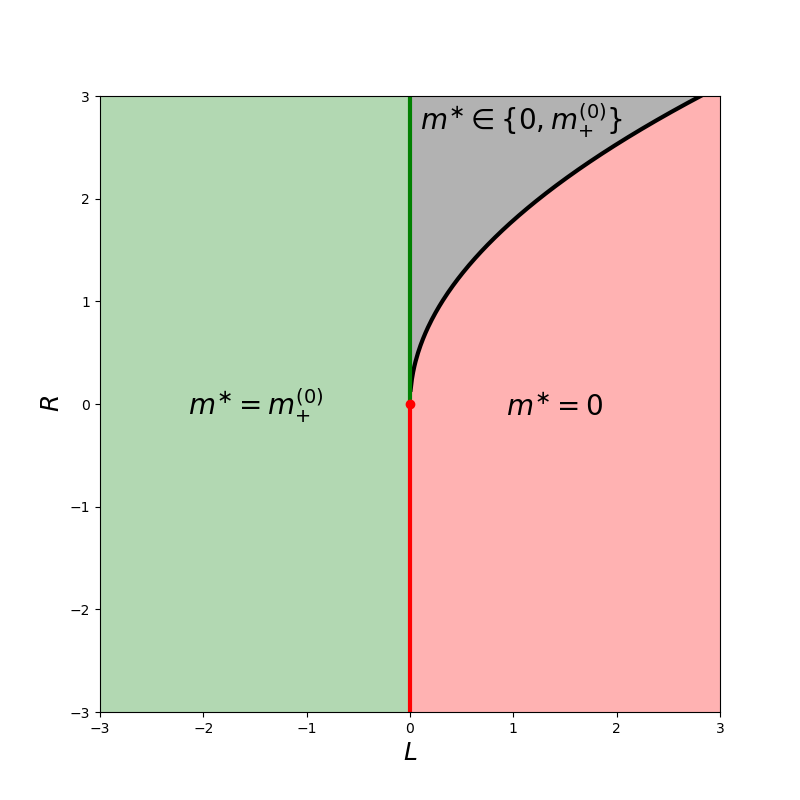}
\caption{Diagram $R\times L$ for case I, where $m^{\ast}=0$ is an absorbing state, while $m=1$ is not.}
\label{diagram_m0}
\end{figure}

In this scenario, one has $\overline{A^{(1)}}=0$, which means that the transition $000\rightarrow 010$ (see Table \ref{t1}) is forbidden, and $A^{(4)}\neq 0$. The equation of motion is cast as $\dot m=F_{0}(m)$, where
\begin{align}
\nonumber F_{0}(m) &:= - Lm + \left(R+2L\right)m^{2} - \left(R+L+A^{(4)}\right)m^{3} \\
 &= -m\left[ L - \left(R+2L\right)m + \left(R+L+A^{(4)}\right)m^{2} \right].
\label{F0}
\end{align}
As seen from \eqref{F0}, apart from the stationary point $m^{\ast}=0$, two other stationary solutions, $m_{+}^{(0)}$ and $m_{-}^{(0)}$ , can emerge depending on the choice of the pair $(L,R)$. These values are given by
\begin{align}
m_{\pm}^{(0)} := \frac{R+2L\pm\sqrt{ R^{2} - 4LA^{(4)} }}{2\left(R+L+A^{(4)}\right)}.
\label{m+-0}
\end{align}
The stationary points in this case are
\begin{align}
m^{\ast}\in\{0,m_{\pm}^{(0)}\}.
\label{mast0}
\end{align}
If $A^{(4)}=0$, the system will also have $m^{\ast}=1$ as one of its stationary points. Since this case ($0$ and $1$ being both fixed points of the dynamical system) will be examined later, it will be assumed that $A^{(4)}\neq 0$ (or $A^{(4)}>0$) in this section. The stability analysis is going to be performed through the (one-dimensional) isocline method, which stems on the analysis of the sign of \eqref{F0}. Consequently, the relative positions of the points $0$, $m_{+}^{(0)}$ and $m_{-}^{(0)}$ plays a major rule in this procedure and is detailed in the Appendix \ref{caseI}, and the results are summarized in figure \ref{diagram_m0}.


Defining the regions
\begin{align}
\left\{
\begin{array}{ll}
\Omega_{0}^{(0)} &:= \left\{(L,R)\in\mathbb{R}^{2}: L\geq0\text{ and }R<\sqrt{4LA^{(4)}}\right\} \cup \left\{(0,0)\right\} \\
\Omega_{+}^{(0)} &:= \left\{(L,R)\in\mathbb{R}^{2}: L<0\right\} \cup \left\{(L,R)\in\mathbb{R}^{2}: L=0\text{ and }R>0\right\} \\
\Omega_{0+}^{(0)} &:= \left\{(L,R)\in\mathbb{R}^{2}: L>0\text{ and }R\geq\sqrt{4LA^{(4)}}\right\}
\end{array}
\right.,
\label{Omega0}
\end{align}
with $\Omega_{0}^{(0)}\cup\Omega_{+}^{(0)}\cup\Omega_{0+}^{(0)}=\mathbb{R}^{2}$. As shown in figure \ref{diagram_m0}, the regions $\Omega_{0}^{(0)}$ and $\Omega_{+}^{(0)}$ contain a single stationary point, which are $m^{\ast}=0$ and $m^{\ast}=m_{+}^{(0)}$, respectively. On the other hand, the region $\Omega_{0+}^{(0)}$ displays two stationary points, $0$ and $m_{+}^{(0)}$; the initial condition $m(t=0)=m_{0}$ determines the trajectory toward one of these two fixed points in $\Omega_{0+}^{(0)}$. In Appendix \ref{caseI}, it is shown that $0<m_{-}^{(0)}<m_{+}^{(0)}<1$ in this region, and
\begin{align}
\lim_{t\rightarrow\infty}m(t) = \left\{
\begin{array}{lcl}
0 &,& m_{0} < m_{-}^{(0)} \\
m_{+}^{(0)} &,& m_{0}>m_{-}^{(0)}
\end{array}
\right..
\label{Omega0+}
\end{align}
In other words, the point $m=m_{-}^{(0)}$, which is an unstable fixed point, is a separatrix between these two stationary points, $0$ and $m_{+}^{(0)}$.

Although the region $\{(L,R):\mathbb{R}^{2}: L=0\text{ and }R<0\}$ belongs to $\Omega_{0}^{(0)}$, the transition to the region $\Omega_{+}^{(0)}$ is smooth in the sense that
\begin{align}
\lim_{ \underset{R\leq 0}{L\rightarrow 0^{-}} }m_{+}^{(0)} = 0
\label{transition0-+}
\end{align}
from \eqref{m+-0}. Conversely,
\begin{align}
\lim_{ \underset{R>0}{L\rightarrow 0^{-}} }m_{+}^{(0)} = \frac{R}{R+A^{(4)}} > 0
\label{transition+-0+}
\end{align}
in the boundary between $\Omega_{+}^{(0)}$ and $\Omega_{0+}^{(0)}$.

In the frontier between $\Omega_{0}^{(0)}$ and $\Omega_{0+}^{(0)}$, one has
\begin{align}
\lim_{ \underset{R>0}{L\rightarrow \left(\nicefrac{R^{2}}{4A^{(4)}}\right)^{-}} }m_{+}^{(0)} = \frac{R}{R+2A^{(4)}}>0,
\label{transition0-0+}
\end{align}
indicating a discontinuity of $m_{+}^{(0)}$ when crossing this boundary. More precisely, the underlying mechanism of this discontinuity is a saddle-node bifurcation \cite{strogatz2024nonlinear}: crossing from the interior of $\Omega_{0}^{(0)}$ to the interior of $\Omega_{0+}^{(0)}$, a stationary point different from $m^{\ast}=0$, given by \eqref{transition0-0+}, emerges in the boundary $R=\sqrt{4LA^{(4)}}$; this point bifurcates into $m_{+}^{(0)}$ (stable) and $m_{-}^{(0)}$ (unstable) in $\Omega_{0+}^{(0)}$.

Summarizing, if one regards $m_{+}^{(0)}$ as an order parameter, the transition is of first-order between $\Omega_{0}^{(0)}$ and $\Omega_{0+}^{(0)}$, continuous between $\Omega_{+}^{(0)}$ and $\Omega_{0}^{(0)}$ or $\Omega_{0+}^{(0)}$. The origin $(L,R)=(0,0)$ is then a tricritical point.


\section{System with absorbing state $m^{\ast}=1$ (without the absorbing state $m^{\ast}=0$): Case II}

\begin{figure}[ht!]
\centering
\includegraphics[width=346pt]{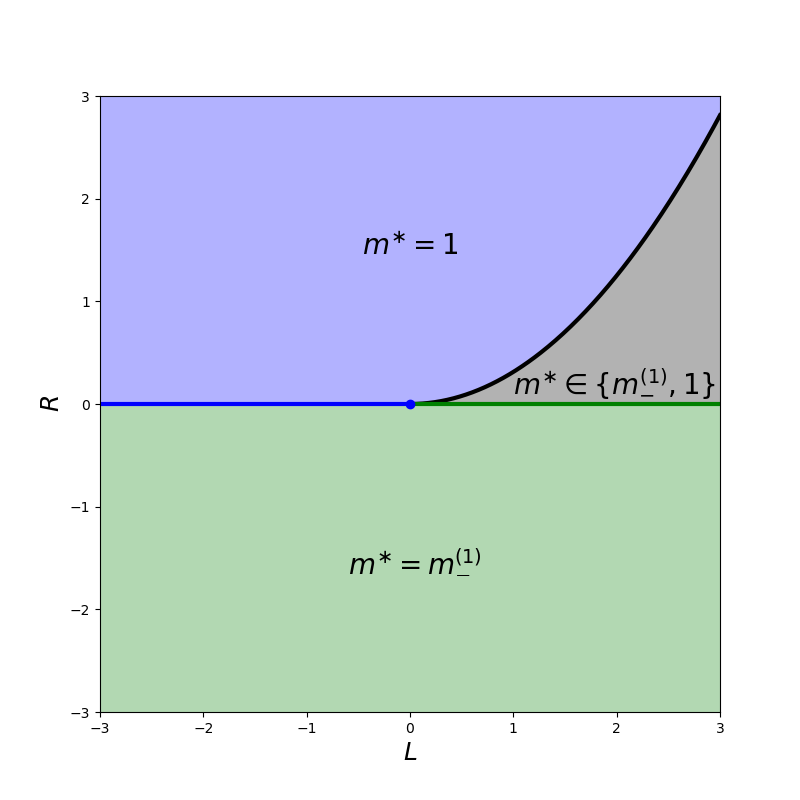}
\caption{Diagram $R\times L$ for case II, where $m^{\ast}=1$ is an absorbing state, while $m=0$ is not.}
\label{diagram_m1}
\end{figure}

In this scenario, one has $A^{(4)}=0$, which forbids the transition $111\rightarrow 101$ (see Table \ref{t1}), and $\overline{A^{(1)}}\neq 0$. The equation of motion is cast as $\dot m=F_{1}(m)$, where
\begin{align}
\nonumber F_{1}(m) &:= \overline{A^{(1)}}-\left(3\overline{A^{(1)}}+L\right)m + \left(3\overline{A^{(1)}} + R + 2L\right)m^{2} - \left(R + L + \overline{A^{(1)}}\right)m^{3} \\
&= \left(1-m\right)\left[ \overline{A^{(1)}} - \left(2\overline{A^{(1)}}+L\right)m + \left(R+L+\overline{A^{(1)}}\right)m^{2}\right]
\label{F1}
\end{align}
Two more stationary points, $m_{+}^{(1)}$ and $m_{-}^{(1)}$, are potentially available, depending on the choice of the parameters, and are given by
\begin{align}
m_{\pm}^{(1)} := \frac{2\overline{A^{(1)}}+L\pm\sqrt{ L^{2} - 4R\overline{A^{(1)}} }}{2\left(R+L+\overline{A^{(1)}}\right)}.
\label{m+-1}
\end{align}
The stationary points in this case are
\begin{align}
m^{\ast}\in\{1,m_{\pm}^{(1)}\}.
\label{mast1}
\end{align}
Analogously in the previous section, the case $\overline{A^{(1)}}=0$ is excluded in the analysis here, since it will be explored in the next section. Moreover, the technical details of the stationary analysis of \ref{F1} is shown in Appendix \ref{caseII}, and the results can be summarized in the diagram of figure \ref{diagram_m1}.

Defining the regions
\begin{align}
\left\{
\begin{array}{ll}
\Omega_{1}^{(1)} &:= \left\{(L,R)\in\mathbb{R}^{2}: R\geq0\text{ and }L<\sqrt{4R\overline{A^{(1)}}}\right\} \cup \left\{(0,0)\right\} \\
\Omega_{-}^{(1)} &:= \left\{(L,R)\in\mathbb{R}^{2}: R<0\right\} \cup \left\{(L,R)\in\mathbb{R}^{2}: R=0\text{ and }L>0\right\} \\
\Omega_{1-}^{(1)} &:= \left\{(L,R)\in\mathbb{R}^{2}: R>0\text{ and }L\geq\sqrt{4R\overline{A^{(1)}}}\right\}
\end{array}
\right.,
\label{Omega1}
\end{align}
with $\Omega_{1}^{(1)}\cup\Omega_{-}^{(1)}\cup\Omega_{1-}^{(1)}=\mathbb{R}^{2}$. As shown in figure \ref{diagram_m1}, the regions $\Omega_{1}^{(1)}$ and $\Omega_{-}^{(1)}$ contain a single stationary point, which are $m^{\ast}=1$ and $m^{\ast}=m_{-}^{(1)}$, respectively. On the other hand, the region $\Omega_{1-}^{(1)}$ displays two stationary points, $1$ and $m_{-}^{(1)}$; the initial condition $m(t=0)=m_{0}$ determines the trajectory toward one of these two fixed points in $\Omega_{1-}^{(1)}$. In Appendix \ref{caseII}, it is shown that $0<m_{-}^{(1)}<m_{+}^{(1)}<1$ in this region, and
\begin{align}
\lim_{t\rightarrow\infty}m(t) = \left\{
\begin{array}{lcl}
1 &,& m_{0} > m_{+}^{(1)} \\
m_{-}^{(0)} &,& m_{0}<m_{+}^{(1)}
\end{array}
\right..
\label{Omega1+}
\end{align}
In other words, the point $m=m_{+}^{(1)}$, which is an unstable fixed point, is a separatrix between these two stationary points, $1$ and $m_{-}^{(1)}$.

Although the region $\{(L,R):\mathbb{R}^{2}: R=0\text{ and }L<0\}$ belongs to $\Omega_{1}^{(1)}$, the transition to the region $\Omega_{-}^{(1)}$ is smooth in the sense that
\begin{align}
\lim_{ \underset{L\leq 0}{R\rightarrow 0^{-}} }m_{-}^{(1)} = 1
\label{transition--1}
\end{align}
from \eqref{m+-1}. Conversely,
\begin{align}
\lim_{ \underset{L>0}{R\rightarrow 0^{-}} }m_{-}^{(1)} = \frac{\overline{A^{(1)}}}{R+\overline{A^{(1)}}} > 0
\label{transition+-1+}
\end{align}
in the boundary between $\Omega_{-}^{(1)}$ and $\Omega_{1-}^{(1)}$.

In the frontier between $\Omega_{1}^{(1)}$ and $\Omega_{1-}^{(1)}$, one has
\begin{align}
\lim_{ \underset{L>0}{R\rightarrow \left(\nicefrac{L^{2}}{4\overline{A^{(1)}}}\right)^{-}} }m_{-}^{(1)} = \frac{2\overline{A^{(1)}}}{2\overline{A^{(1)}}+L}>0,
\label{transition1-1-}
\end{align}
indicating a discontinuity of $m_{-}^{(1)}$ when crossing this boundary. More precisely, the underlying mechanism of this discontinuity is a saddle-node bifurcation \cite{strogatz2024nonlinear}: Crossing from the interior of $\Omega_{1}^{(1)}$ to the interior of $\Omega_{1-}^{(1)}$, a stationary point different from $m^{\ast}=1$, given by \eqref{transition1-1-}, emerges in the boundary $L=\sqrt{4R\overline{A^{(1)}}}$; this point bifurcates into $m_{-}^{(1)}$ (stable) and $m_{+}^{(1)}$ (unstable) in $\Omega_{1-}^{(1)}$.

Summarizing, if one regards $m_{-}^{(1)}$ as an order parameter, the transition is of first-order between $\Omega_{1}^{(1)}$ and $\Omega_{1-}^{(1)}$, continuous between $\Omega_{-}^{(1)}$ and $\Omega_{1}^{(1)}$ or $\Omega_{1-}^{(1)}$. The origin $(L,R)=(0,0)$ is then a tricritical point.


\section{System with absorbing state $m^{\ast}=0$ and $m^{\ast}=1$: Case III}

In this last scenario, one has $A^{(4)}=0$ and $\overline{A^{(1)}}=0$, which forbid the transitions $111\rightarrow 101$ and $000\rightarrow 010$ (see Table \ref{t1}). The equation of motion is cast as $\dot m=F_{01}(m)$, where
\begin{align}
\nonumber F_{01}(m) &:= - Lm + \left(R + 2L\right)m^{2} - \left(R + L\right)m^{3} \\
 &= m\left(1-m\right)\left[\left(R+L\right)m-L\right],
\label{F01}
\end{align}
and the stationary points $m^{\ast}=0$ and $m^{\ast}=1$ are naturally present. Apart from them, it is also possible to have a third one,
\begin{align}
\tilde{m}^{(01)} := \frac{L}{R+L},
\label{tildem}
\end{align}
depending on the choice of the pair $(L,R)$. The set of stationary points in this case are, therefore,
\begin{align}
m^{\ast}\in\{0,1,\tilde{m}^{(01)}\}.
\label{mast01}
\end{align}
The stability analysis is summarized in figure \ref{diagram_m01}, while the technical details are found in Appendix \ref{caseIII}.

\begin{figure}[ht!]
\centering
\includegraphics[width=346pt]{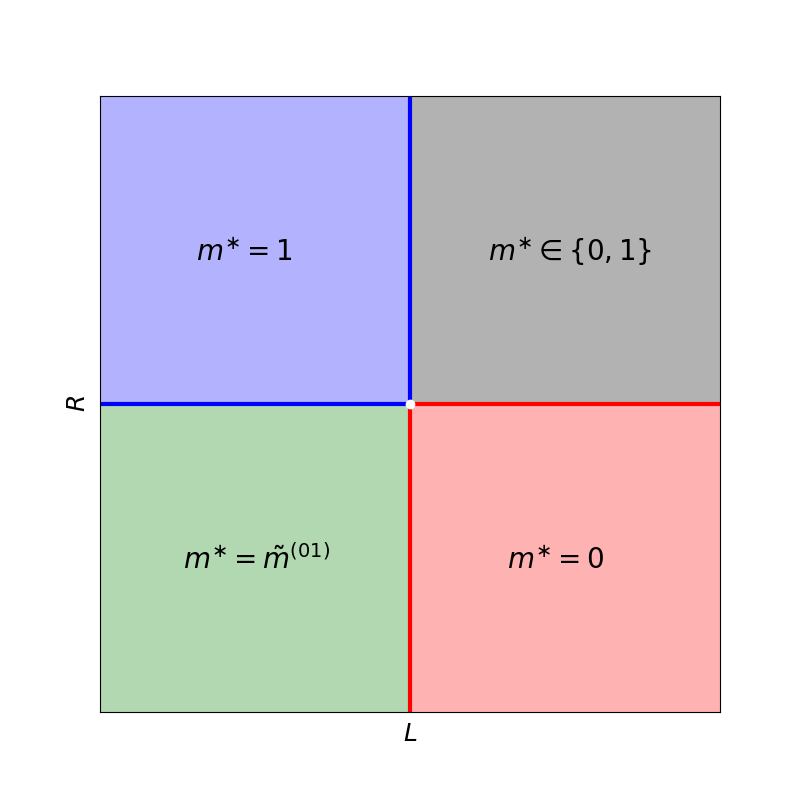}
\caption{Diagram $R\times L$ for case III, where $m^{\ast}=0$ and $m^{\ast}=1$ are absorbing states.}
\label{diagram_m01}
\end{figure}

Defining the regions
\begin{align}
\left\{
\begin{array}{ll}
\Omega_{0}^{(01)} &:= \left\{(L,R)\in\mathbb{R}^{2}: L\geq 0\text{ and }R\leq 0\right\} \setminus \{(0,0)\} \\
\Omega_{1}^{(01)} &:= \left\{(L,R)\in\mathbb{R}^{2}: L\leq 0\text{ and }R\geq 0\right\} \setminus \{(0,0)\} \\
\Omega_{\tilde m}^{(01)} &:= \left\{(L,R)\in\mathbb{R}^{2}: L<0\text{ and }R<0\right\} \\
\Omega_{01}^{(01)} &:= \left\{(L,R)\in\mathbb{R}^{2}: L>0\text{ and }R>0\right\}
\end{array}
\right.,
\label{Omega01}
\end{align}
with $\Omega_{0}^{(01)}\cup\Omega_{1}^{(01)}\cup\Omega_{\tilde m}^{(01)}\cup\Omega_{01}^{(01)}=\mathbb{R}^{2}\setminus\{(0,0)\}$. As shown in figure \ref{diagram_m01}, the regions $\Omega_{0}^{(01)}$, $\Omega_{1}^{(01)}$ and $\Omega_{\tilde m}^{(01)}$ contain a single stationary point, which are $m^{\ast}=0$, $m^{\ast}=1$ and $m^{\ast}=\tilde{m}^{(01)}$, respectively. On the other hand, the region $\Omega_{01}^{(01)}$ displays two stationary points, $0$ and $1$; the initial condition $m(t=0)=m_{0}$ determines the trajectory toward one of these two fixed points in $\Omega_{01}^{(01)}$. In Appendix \ref{caseIII}, it is shown that $0<\tilde{m}^{(01)}<1$ in this region, and
\begin{align}
\lim_{t\rightarrow\infty}m(t) = \left\{
\begin{array}{lcl}
0 &,& m_{0} < \tilde{m}^{(01)} \\
1 &,& m_{0} > \tilde{m}^{(01)}
\end{array}
\right..
\label{Omega01+}
\end{align}
In other words, the point $m=\tilde{m}^{(01)}$, which is an unstable fixed point, is a separatrix between these two stationary points, $0$ and $1$.

Although the region $\{(L,R):\mathbb{R}^{2}: R=0\text{ and }L<0\}$ belongs to $\Omega_{1}^{(01)}$, and $\{(L,R):\mathbb{R}^{2}: L=0\text{ and }R<0\}$ belongs to $\Omega_{0}^{(01)}$, the transition between $\Omega_{\tilde m}^{(01)}$ and these involved regions are smooth in the sense that
\begin{align}
\lim_{ \underset{L<0}{R\rightarrow 0^{-}} }\tilde{m}^{(01)} = 1 \quad\textnormal{ and }\quad \lim_{ \underset{R<0}{L\rightarrow 0^{-}} }\tilde{m}^{(01)} = 0.
\label{trnasitions-smooth01}
\end{align}
from \eqref{tildem}.

The stationary behvior of the point $(0,0)$ should be examined separately, since the equation of motion is $\dot m=0$ in this point when $A^{(4)}=\overline{A^{(1)}}=0$. This means that no dynamics takes place at this particular point and the stationary value for $m(t)$ equals the initial value.


\section{The role of the master node}

There are some general observations concerning the variables
\begin{align}
L := A^{(2)} - \overline{B^{(1)}} - \overline{B^{(3)}} \quad\textnormal{ and }\quad R := \overline{A^{(3)}} - B^{(2)} - B^{(4)},
\end{align}
where the involved quantities (defined before) are stated here for clarity in table \ref{ABBABB}.
\begin{table}[ht!]
\centering
\begin{tabular}{ccccccc}
$010$&$\overset{A^{(2)}}{\longrightarrow}$&$000$ & \hspace{50pt} & $101$&$\overset{\overline{A^{(3)}}}{\longrightarrow}$&$111$ \\
 & & & & & & \\
$001$&$\overset{\overline{B^{(1)}}}{\longrightarrow}$&$011$ & \hspace{50pt} & $011$&$\overset{B^{(2)}}{\longrightarrow}$&$001$ \\
 & & & & & & \\
$100$&$\overset{\overline{B^{(3)}}}{\longrightarrow}$&$110$ & \hspace{50pt} & $110$&$\overset{B^{(4)}}{\longrightarrow}$&$100$
\end{tabular}
\caption{Transition rates $A^{(2)}$, $\overline{B^{(1)}}$, $\overline{B^{(3)}}$, $\overline{A^{(3)}}$, $B^{(2)}$ and $B^{(4)}$.}
\label{ABBABB}
\end{table}

Some general observations can then be made from these transition rates.
\begin{enumerate}

\item Tendency to state $m=0$: $L>0$ and $R<0$, as a consequence of large $A^{(2)}$, $B^{(2)}$ and $B^{(4)}$, and low $\overline{A^{(3)}}$, $\overline{B^{(1)}}$ and $\overline{B^{(3)}}$. This is valid even $m=0$ not being an absorbing state.

\item Tendency to state $m=1$: $L<0$ and $R>0$, as a consequence of low $A^{(2)}$, $B^{(2)}$ and $B^{(4)}$, and large $\overline{A^{(3)}}$, $\overline{B^{(1)}}$ and $\overline{B^{(3)}}$. This is valid even $m=1$ not being an absorbing state.

\item Tendency to cluster formation: $L>0$ and $R>0$, as a consequence of large $A^{(2)}$ and $\overline{A^{(3)}}$.

\item Tendency to cluster destruction: $L<0$ and $R<0$, as a consequence of low $A^{(2)}$ and $\overline{A^{(3)}}$.
  
\end{enumerate}
  
Note that these four cases are not disjoint, but one can see the potential impact of the master node in the dynamics. To facilitate the analysis, let us consider two extremal cases. First, if the master node is decoupled by letting $\Gamma_{i}=0$ for all sites in \eqref{AB}, it leads to
\begin{align}
L(\Gamma_{i}=0) = p_{2} - \overline{w_{1}} - \overline{w_{3}} \quad\textnormal{ and }\quad R(\Gamma_{i}=0) = \overline{p_{3}} - w_{2} - w_{4}.
\label{Gamma0}
\end{align}
On the other hand, the other extremal case, $\Gamma_{i}=1$ for all vertices $i$, leads to
\begin{align}
L(\Gamma_{i}=1) = q_{2} - \overline{v_{1}} - \overline{v_{3}} \quad\textnormal{ and }\quad R(\Gamma_{i}=1) = \overline{q_{3}} - v_{2} - v_{4}.
\label{Gamma1}
\end{align}
From \eqref{Gamma0} and \eqref{Gamma1}, all the involved transition rates are independent. This means that starting from a setup where the master node has no role ($\Gamma_{i}=0$), one can move to another scenario by letting the master node participate in the dynamics. As an example, one might choose the parameters such that the dynamics tends to the state $m=0$, but turning on the master node leads the system to the state $m=1$.

\indent

\section{Discussion}

\textbf{Model analysis and key findings.} The master-node network model aims to provide a broad and unifying framework for the study of opinion dynamics, much like other general approaches in sociophysics \cite{martins12b,Boettcher2017,Galam2020}. The mean-field analysis presented here reveals that a large portion of the parameter space can be partitioned into three distinct absorbing-state regimes. By appropriately choosing the parameters, the model is flexible enough to reproduce classical mechanisms in opinion dynamics, such as the presence of contrarians or inflexibles. The adoption of the annealing approximation, together with mean-field techniques, transforms the otherwise complex many-body stochastic process into a tractable analytical framework while still retaining the essential ingredients of the dynamics. The main theoretical advance of this work lies in the complete characterization of phase diagrams for all three cases, covering convergence to complete disagreement ($m^*=0$), convergence to complete consensus ($m^*=1$), and bistability where the outcome depends on the initial condition of the system. 

This study builds directly on the master-node network model previously introduced in \cite{ferreira2020stochastic,mihara2023critical}. In those works, agents were arranged in a one-dimensional ring with nearest-neighbor interactions, and a master node exerted external influence over a quenched fraction $\rho$ of agents. The master node imposed its state with strength $r$, while $\Delta$ quantified the intrinsic bias of agents toward selfish ($s=0$) or cooperative ($s=1$) behavior. Across these formulations, the central object of interest was the absorbing-state phase transition, measured through the order parameter $I$ (or equivalently $m$), which describes the fraction of agents opposing the master node’s state.

Compared with \cite{ferreira2020stochastic}, the present work provides a substantially more general formulation by introducing chirality via the parameters $L$ and $R$, which quantify left- and right-oriented asymmetries in influence. This extension greatly enriches the dynamical landscape, giving rise to both continuous and discontinuous transitions and unveiling the existence of tricritical organization around $(L,R)=(0,0)$. Another methodological advance is the adoption of the annealing approximation, in which the quenched disorder represented by $\rho$ is replaced by its mean value. This simplification makes the mean-field analysis analytically tractable and allows us to map out complete phase diagrams, classifying the system into three absorbing-state regimes. Consequently, the present model yields diagrams that are richer and more informative than those reported in \cite{ferreira2020stochastic}, which primarily studied continuous transitions through Monte Carlo simulations and finite-size scaling techniques.

Relative to \cite{mihara2023critical}, which emphasized critical exponents obtained from extensive Monte Carlo simulations, the present study represents a complementary but equally important contribution. Although that earlier work demonstrated that the model does not belong to universality classes such as directed percolation or the voter model, here we adopt an analytical approach based on annealing approximation and mean-field theory. The inclusion of chirality through $L$ and $R$ allows for explicit equations of motion and the construction of complete phase diagrams. This broader formulation highlights how asymmetric interactions reshape consensus and polarization, generating dynamical regimes that could not be captured by the original three-parameter version $(r,\rho,\Delta)$.

Taken together, the three studies demonstrate continuity and progression. While all share a common structural foundation and focus on absorbing-state transitions in networks with a master node, the present contribution advances the field by generalizing the model to include chirality, developing an analytical framework that unifies continuous and discontinuous transitions, and uncovering tricritical organization. This places our results as a theoretical extension that consolidates and surpasses the findings of \cite{ferreira2020stochastic,mihara2023critical}.

\indent
\textbf{Phase transitions and critical behavior.} In order to clarify the critical behavior, let us examine separately the scenarios with a single trivial stationary point (Cases I and II) and the case with two trivial absorbing states (Case III). In the former, three dynamical regions can be identified: $\Omega_{\text{trivial}}$, where only the trivial fixed point exists ($m^*=0$ in Case I, $m^*=1$ in Case II); $\Omega_{\text{nontrivial}}$, where a new nontrivial stationary point emerges; and $\Omega_{2}$, where the system becomes bistable with coexistence of trivial and nontrivial solutions. The transition from $\Omega_{\text{trivial}}$ to $\Omega_{\text{nontrivial}}$ is continuous, in line with the behavior described in \eqref{transition0-+} and \eqref{transition--1}. By contrast, the transition from $\Omega_{\text{nontrivial}}$ to $\Omega_{2}$ is discontinuous and associated with a saddle-node bifurcation. 

It is essential to emphasize that this saddle-node bifurcation  take place along the boundary curves $R = \sqrt{4LA^{(4)}}$ (Case I) and $L = \sqrt{4R\overline{A^{(1)}}}$ (Case II). These boundaries separate the pink (monostable) and gray (bistable) regions in Fig.~2. At the boundary, a single nontrivial fixed point emerges and then splits into two distinct solutions inside the gray region: a stable node ($m_{+}$) and an unstable saddle ($m_{-}$). Crucially, the trivial solution ($m^*=0$ in Case I, or $m^*=1$ in Case II) remains present in both regions and is unaffected by the bifurcation.

The region $\Omega_{2}$ therefore contains two attractors, and the initial condition plays a decisive role in determining which stationary state the dynamics converge to. The unstable solution $m_{-}$ functions as a separatrix between the basins of attraction. The coexistence of two stable solutions ($m^*=0$ and $m_{+}$, or $m^*=1$ and $m_{+}$ depending on the case) reflects a first-order transition mediated by a saddle-node bifurcation. 

In Case III, the situation differs in that both $m^*=0$ and $m^*=1$ are absorbing states. The corresponding domains, $\Omega_{0}^{(01)}$ and $\Omega_{1}^{(01)}$, are closed, in the sense that the stationary value of $m(t)$ matches the absorbing state. At the origin $(0,0)$ the dynamics are degenerate: The stationary point simply equals the initial condition. The mixed region $\Omega_{01}^{(01)}$ thus behaves analogously to $\Omega_{2}$ in Cases I and II, in that the system’s fate depends on initial conditions, but without the appearance of a discontinuous transition or a saddle-node bifurcation. Instead, $(0,0)$ plays a structural role by defining the polarization threshold $\tilde{m}^{(01)} = L/(R+L)$.  

Altogether, these results provide a coherent picture: $(0,0)$ acts as a tricritical organizer, structuring the global phase diagram, while the actual saddle-node bifurcations that generate bistability occur along finite-$L$/$R$ boundaries. This distinction clarifies the relationship between continuous and discontinuous transitions and shows how chirality governs the onset of bistability in the model. The parameters $L$ and $R$ thus determine not only the final state of consensus or polarization but also the very nature of the critical behavior—continuous, discontinuous, or tricritical.

By situating these results alongside previous work, we conclude that the inclusion of chirality and the use of annealing approximation not only generalize the master-node model but also reveal fundamentally new critical properties absent in the quenched-disorder formulations. This analytical framework deepens our understanding of how asymmetric influence can drive consensus, polarization, and bistability in social systems.


\subsection{Interdisciplinary implications}

Although our model is formulated in the language of mathematics and inspired by concepts from statistical mechanics, it aims to describe profoundly social phenomena: how asymmetric influences, network structures, and initial conditions shape collective opinion dynamics—ultimately leading to consensus, polarization, or abrupt collapse.

Abrupt transitions in opinion dynamics have sparked debate, as certain models yield qualitatively different predictions depending on how the thermodynamic limit is taken \cite{Sznajd-Weron_2011, Galam_2011, PhysRevE.105.064306}. Our work resolves this ambiguity by providing a mathematically robust framework in which discontinuous transitions emerge naturally from the underlying dynamics of chiral asymmetry—rather than as artifacts of computational or limiting procedures. While grounded in the formalism of nonequilibrium statistical mechanics and phase transitions, the model’s central insights—particularly regarding asymmetric influence, tipping points, bistability, and consensus collapse—resonate strongly with long-standing questions in sociology, political science, and social psychology.

The notion of a \emph{master node} exerting asymmetric, chiral influence aligns closely with sociological theories of \emph{opinion leadership} and influencer dynamics in social networks \cite{DONG2017187, Zhaoetal18}. This conceptual parallel extends to Katz and Lazarsfeld’s seminal \emph{two-step flow of communication} model \cite{katz2017personal}, historically developed (with Edward Shils) to challenge the simplistic “hypodermic needle” theory of media effects \cite{pooley2006}. Like our master node, opinion leaders in this framework act as pivotal intermediaries who channel, filter, and amplify external influence within social networks.

Directional bias—or chirality—in our model captures asymmetric influence, which arises naturally in contexts featuring clear hierarchies, such as those between opinion leaders and followers \cite{DONG2017187,Zhaoetal18}, digital influencers \cite{helfmann2023}, or when agents exhibit differential responsiveness to competing options \cite{e22010025}. This asymmetry is central to understanding how small imbalances can drive large-scale outcomes.

Our findings on discontinuous transitions and bistability echo Granovetter’s threshold models of collective behavior \cite{granovetter1978}, which show that macro level shifts—such as protest cascades or market tipping points—depend critically on the \emph{distribution} of individual thresholds, not merely their average. Similarly, in our framework, system-level outcomes hinge on the fine-grained structure of influence and initial conditions, rather than on aggregate statistical moments.

The identification of tricritical points and bistable regimes—where systems exhibit path dependence and resistance to perturbations—finds direct parallels in political science literature on policy feedback and institutional lock-in \cite{pierson2000}. In both cases, once a system settles into an absorbing state, minor shocks are insufficient to reverse its trajectory; only structural interventions (in our model, changes in the parameters \(L\) and \(R\)) can shift the equilibrium.

Moreover, the role of directional bias in driving polarization aligns with Sunstein’s account of echo chambers and group polarization \cite{sunstein2009}. In such settings, homophilic interactions amplify existing views not through manipulation, but via ordinary psychological mechanisms—social validation and reinforcement—mirroring the chiral feedback loops in our model.

Finally, the sensitivity of consensus outcomes to initial conditions—especially in bistable regimes—reflects well-established findings in social psychology on priming and framing effects \cite{druckman2001}. Druckman demonstrates that framing influences preferences not automatically, but through mediating factors such as source credibility, individual predispositions, and deliberation. Our model incorporates this insight by treating the initial magnetization \(m_0\) as a decisive variable that shapes the system’s long-term trajectory.

These mechanisms are especially relevant for modeling—and understanding—theoretical aspects of real-world social systems where a centralized source of influence (e.g., media outlets, political leaders, digital platforms, or parties) exerts asymmetric pressure on a distributed network of agents. In such contexts, collective outcomes often hinge critically on initial opinion configurations and the precise architecture of influence.

The framework also applies to scenarios where minor structural changes in influence networks trigger qualitative behavioral shifts, or where systems remain locked into specific opinion states even after the original source of influence is removed—phenomena described in the literature as path dependence in public policy \cite{pierson2000} and persistent polarization in echo chambers \cite{sunstein2009}.

Concrete examples include the following:

\begin{itemize}
\item The spread of misinformation or political messaging on social media, where recommendation algorithms act as master nodes that asymmetrically promote content, potentially leading to abrupt consensus or deep polarization \cite{sunstein2009, granovetter1978};
    \item Social mobilization and protest dynamics, where small groups of instigators succeed or fail based on the distribution of individual participation thresholds \cite{granovetter1978};
    \item Institutional inertia in public policy, where systems remain entrenched in suboptimal equilibria not due to active opposition, but because of path-dependent lock-in \cite{pierson2000};
    \item Public health campaigns, whose effectiveness depends not only on message content but also on source credibility and how the population is initially framed \cite{druckman2001, meyerowitz1987};
    \item Market dynamics and technology adoption, which often exhibit abrupt transitions from niche to mainstream once a critical mass is reached \cite{granovetter1978, rogers1975}.
\end{itemize}

The interdisciplinary insights developed above—linking chiral asymmetry, initial-condition sensitivity, and abrupt transitions to sociological theories of influence, institutional lock-in, and echo chambers—find their most precise theoretical counterpart in the recent classification by Van Santen et al. \cite{van2024social}, who established a foundational framework for opinion dynamics under centralized external influence. Their model distinguishes between symmetric and asymmetric interaction strategies, demonstrating how a master like agent can drive consensus either toward or against its own opinion, depending on network structure and clustering effects. This aligns directly with our core premise: that a single, asymmetrically connected node can fundamentally reshape collective outcomes by breaking the symmetry of peer interactions.
In this broader context, our chiral master-node model both confirms and transcends Van Santen \textit{et al}.'s paradigm \cite{van2024social}. While they identify macroscopic regimes—consensus, disorder, or anticonsensus — as continuous functions of an external field strength $E_A$, we reveal that under chiral asymmetry ($L \neq R$), the system undergoes qualitative phase transitions governed not by gradual drift, but by structural instabilities. The emergence of a saddle-node bifurcation along the boundary $L^2 = 4R\overline{A^{(1)}}$ (Case II) or $R^2 = 4LA^{(4)}$ (Case I) generates discontinuous jumps between stable states, a phenomenon absent in their continuous framework.
Crucially, our work introduces two interrelated dimensions absent in Van Santen \textit{et al}.'s classification \cite{van2024social}: (i) the existence of a tricritical point at $(L,R)=(0,0)$, which organizes the entire phase diagram by marking the termination of continuous transitions and the birth of discontinuous ones, and (ii) bistability arising from the geometric separation of stable fixed points ($m_+$ and $m_-$) along the boundary between monostable and multistable regions, as visualized in Fig. 2. These features are not mere quantitative extensions—they represent a qualitative shift in the nature of influence: from a tunable perturbation to a topological reorganization of the state space.
Where Van Santen \textit{et al}. \cite{van2024social} describe how social clustering amplifies or resists influence, we show how directional bias—encoded in the chiral parameters $L$ and $R$—determines whether the system evolves smoothly toward consensus or fractures abruptly into polarization. In their model, the transition from disagreement to consensus is monotonic; in ours, it is geometrically constrained by the emergence of unstable separatrices, making outcomes exquisitely sensitive to initial conditions even in the absence of memory, inflexible agents, or heterogeneity. The point $(0,0)$ is not merely a parameter value—it is the apex of a bifurcation landscape where the very character of collective behavior changes.
Thus, while Van Santen et al.\cite{van2024social} provide a powerful taxonomy of opinion regimes under centralized influence, our model unveils the underlying dynamical architecture: a phase space organized by chirality, punctuated by a tricritical singularity, and governed by irreversible, nonmonotonic transitions. We do not replace their framework—we embed it within a richer mathematical structure, revealing that the most profound societal shifts arise not from stronger signals, but from the geometry of asymmetry itself.


\section{Conclusion}

This study presents a significant contribution to the understanding of opinion dynamics within complex systems through a master node model under a mean-field annealing approximation framework. By introducing a chiral formulation of the master node, we were able to uncover a rich phase diagram that includes both continuous and discontinuous transitions between ordered (full consensus or complete disagreement) and disordered (coexistence of opinions) states, where the discontinuity is triggered by a saddle-node bifurcation. The chiral parameters $R$ and $L$ emerge as powerful descriptors of system behavior, quantifying the inherent asymmetry in opinion propagation through directional influences. They determine the ultimate consensus state and helps explain polarization phenomena. We also identified regions in the phase diagram where these transitions are highly dependent on initial conditions. The identification of tricritical points at $(L, R) = (0, 0)$ further highlights the limitations in smoothly connecting disordered and ordered phases, reinforcing the idea that consensus formation is not always a gradual process and can exhibit abrupt shifts under certain conditions. The detailed analysis of transition lines offer practical tools for predicting when small changes in network parameters may lead to dramatic shifts in collective opinion.

One of the key theoretical contributions lies in the analytical framework developed, which allows for the characterization of symmetry-breaking effects induced by asymmetric influence. This provides valuable insights into how centralized nodes shape collective outcomes, paving the way for applications in organizational management, public policy, and misinformation control. The stability analysis performed using isocline methods revealed the emergence of a third stationary point $ \tilde{m}^{(01)} = \frac{L}{R + L} $, depending on the region of the $(L, R)$ parameter space. This finding enriches our understanding of the role of asymmetry in shaping the final state of the system, acting as a biasing mechanism that can either accelerate or delay consensus depending on the coupling parameters.

Our starting point was a general model with several free parameters. Despite this complexity, its mean-field formulation, combined with an annealing approximation in the way the master node influences the spins, led to an unexpectedly simplified description governed by a specific combination of the original parameters, ultimately reducing them to four key quantities ($R$, $L$, $A^{(4)}$, and $\overline{A^{(1)}}$), which could be investigated analytically. While this constitutes our main contribution, it also represents a limitation, as the resulting framework deviates from the original model.

We hope this study sheds light on the fundamental mechanisms governing opinion dynamics in structured populations, particularly in systems where central nodes exert asymmetric influence over collective outcomes. By framing opinion formation through the lens of statistical mechanics and phase transitions, we bridge the gap between abstract modeling and practical applications in sociophysics. Looking ahead, future research should aim to extend the current mean-field approximation to include spatial correlations and heterogeneous network structures, bringing the model closer to empirical observations. Moreover, developing methods to measure the $R$ and $L$ parameters in real-world social networks could enable more accurate predictions and interventions in areas such as marketing, political campaigns, and crisis communication.

Ultimately, this line of inquiry contributes to a deeper understanding of how information, influence, and consensus evolve in complex adaptive systems — a question of growing importance in our increasingly interconnected world. The master node's role as a symmetry-breaking element suggests design principles for influencing real-world social systems, from organizational management to public opinion shaping.

Future research directions should focus on: (i) extending the model beyond mean-field approximation to incorporate spatial correlations, (ii) developing methods to empirically measure the $R$ and $L$ parameters in real social networks, and (iii) exploring applications in controlling misinformation spread and engineering consensus in complex organizations. The mathematical framework developed here provides a solid foundation for these important next steps in sociophysics research.

\section{Acknowledgements}
FFF thanks CNPq for support, Grant No. 316664/2021-9.

\renewcommand{\thesection}{\Alph{section}}
\renewcommand{\theequation}{\thesection.\arabic{equation}}
\setcounter{section}{0}
\setcounter{equation}{0}

\section{Appendix: case $\overline{A^{(1)}}=0$ and $A^{(4)}>0$}
\label{caseI}

As mentioned in the main text, the stability analysis of the stationary points are based on the (one-dimensional) isocline method. This means that one investigates the sign of $F_{0}$ in the differential equation $\dot m=F_{0}(m)$, which means that the relative positions of the key points $0$, $1$, $m_{-}^{(0)}$ and $m_{+}^{(0)}$ are helpful. Although $m=1$ is not a stationary point, it is included in the analysis since the meaningful interval for $m$ is $[0,1]$. The analysis of $m_{-}^{(0)}$ and $m_{+}^{(0)}$ will be shown for reference even if they do not belong to this meaningful interval, though the physical outcome is restricted to the stability of these key points in $[0,1]$ only. The analysis is divided into some parts, as specified in table \ref{parts0}.

\begin{table}[ht!]
\centering
\begin{tabular}{|c|l|}\hline
\multirow{3}{*}{Part I} & Point $(0,0)$ \\\cline{2-2}
 & Point $(A^{(4)}, -2A^{(4)})$ \\\cline{2-2}
 & Point $(0, -A^{(4)})$ \\\hline
\multirow{3}{*}{Part II} & Line $R+L+A^{(4)}=0$ \\\cline{2-2}
 & Line $L=0$ \\\cline{2-2}
 & Curve $R^{2}-4LA^{(4)}=0$ \\\hline
\multirow{6}{*}{Part III} & Region (0) \\\cline{2-2}
 & Region (I) \\\cline{2-2}
 & Region (II) \\\cline{2-2}
 & Region (III) \\\cline{2-2}
 & Region (IV) \\\cline{2-2}
 & Region (V) \\\cline{2-2}
 & Region (VI) \\\hline
\end{tabular}
\caption{Division of the $(L, R)$ plane suitable for the construction of the flux diagram of the case $\overline{A^{(1)}}=0$ and $A^{(4)}>0$; the regions in Part III are defined in picture \ref{regions0}.}
\label{parts0}
\end{table}


\subsection{Part I (points)} 

Three situations are investigated in this part.

\begin{itemize}

\item \underline{I.1) Point $(0,0)$}

\bigskip
The differential equation for this case is
\begin{align}
\dot m = -A^{(4)}m^{3},
\label{I1_1}
\end{align}
and displays a single stationary point at $m^{\ast}=0$.

\item \underline{I.2) Point $(A^{(4)}, -2A^{(4)})$}

\bigskip
The differential equation for this case is
\begin{align}
\dot m = -A^{(4)}m,
\label{I2_1}
\end{align}
and displays a single stationary point at $m^{\ast}=0$.

\item \underline{I.3) Point $(0,-A^{(4)})$}

\bigskip
The differential equation for this case is
\begin{align}
\dot m = -A^{(4)}m^{2},
\label{I3_1}
\end{align}
and displays a single stationary point at $m^{\ast}=0$.

\end{itemize}

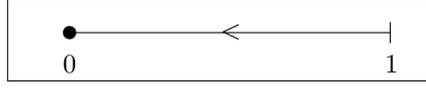
\begin{figure}[ht]
\centering
\fbox{\begin{picture}(150, 25)  
\put(20, 15){\line(1,0){120}}
\put(20, 15){\circle*{5}} \put(140, 11){\line(0, 1){8}}
\put(76.5, 12.75){$<$}
\put(17.5, 0){$0$} \put(138, 0){$1$}
\end{picture}
}
\caption{Flux diagram corresponding to the cases I.1, I.2 and I.3.}
\label{diagram_m0_point}
\end{figure}

The three cases above share the same flux diagram, which is depicted in figure \ref{diagram_m0_point}.

\subsection{Part II (curves)} 

The stability analysis now are performed on lines and curves.

\bigskip
\noindent
\underline{II.1) Line $R+L+A^{(4)}=0$}

\bigskip
The differential equation for this case is
\begin{align}
\dot m = \left(L-A^{(4)}\right)m\left(m-\tilde m\right),
\label{II1_1}
\end{align}
where $\tilde m:=\frac{L}{L-A^{(4)}}$. The system displays two stationary points at $m^{\ast}=0$ and $m^{\ast}=\tilde m$. Since
\begin{align}
\tilde m :=\frac{L}{L-A^{(4)}} \left\{
\begin{array}{llcl}
>1        & \text{ (meaning $0<1<\tilde m$) } &,& L>A^{(4)} \\
\in (0,1) & \text{ (meaning $0<\tilde m<1$) } &,& L<0 \\
<0        & \text{ (meaning $\tilde m<0<1$) } &,& 0<L<A^{(4)}
\end{array}
\right.,
\label{II1_2}
\end{align}
the flux diagram is shown in figure \ref{partII1}.

\begin{figure}[ht]
\centering
\fbox{\begin{picture}(150, 50)  
\put(20, 40){\fbox{$L>A^{(4)}$}}
\put(20, 15){\line(1,0){60}} \multiput(80, 15)(2, 0){30}{\line(1, 0){1}}
\put(20, 15){\circle*{5}} \put(80, 11){\line(0, 1){8}} \put(140, 15){\circle*{5}}
\put(46.5, 12.75){$<$} \put(106.5, 12.75){$<$}
\put(17.5, 0){$0$} \put(78, 0){$1$} \put(136, 0){$\tilde m$}
\end{picture}
}
\fbox{\begin{picture}(150, 50)  
\put(20, 40){\fbox{$L<0$}}
\put(20, 15){\line(1,0){120}}
\put(20, 15){\circle*{5}} \put(80, 15){\circle*{5}} \put(140, 11){\line(0, 1){8}}
\put(46.5, 12.75){$>$} \put(106.5, 12.75){$<$}
\put(17.5, 0){$0$} \put(76, 0){$\tilde m$} \put(138, 0){$1$}
\end{picture}
}
\fbox{\begin{picture}(150, 50)  
\put(20, 40){\fbox{$0<L<A^{(4)}$}}
\multiput(20, 15)(2, 0){30}{\line(1, 0){1}} \put(80, 15){\line(1,0){60}}
\put(20, 15){\circle*{5}} \put(80, 15){\circle*{5}} \put(140, 11){\line(0, 1){8}}
\put(46.5, 12.75){$>$} \put(106.5, 12.75){$<$}
\put(16, 0){$\tilde m$} \put(77.5, 0){$0$} \put(138, 0){$1$}
\end{picture}
}
\caption{Flux diagram - part II.1. The dots represent stationary points and the dashed lines indicate regions outside $[0,1]$.}
\label{partII1}
\end{figure}

This case assumes $L\neq A^{(4)}$ and $L\neq 0$; nevertheless, these two situations were analyzed previously (part I.2 and I.3, respectively).

\bigskip
\noindent
\underline{II.2) Line $L=0$}

\begin{figure}[ht]
\centering
\fbox{\begin{picture}(150, 50)  
\put(20, 40){\fbox{$R<-A^{(4)}$}}
\put(20, 15){\line(1,0){60}} \multiput(80, 15)(2, 0){30}{\line(1, 0){1}}
\put(20, 15){\circle*{5}} \put(80, 11){\line(0, 1){8}} \put(140, 15){\circle*{5}}
\put(46.5, 12.75){$<$} \put(106.5, 12.75){$<$}
\put(17.5, 0){$0$} \put(78, 0){$1$} \put(136, 0){$\tilde m$}
\end{picture}
}
\fbox{\begin{picture}(150, 50)  
\put(20, 40){\fbox{$R>0$}}
\put(20, 15){\line(1,0){120}}
\put(20, 15){\circle*{5}} \put(80, 15){\circle*{5}} \put(140, 11){\line(0, 1){8}}
\put(46.5, 12.75){$>$} \put(106.5, 12.75){$<$}
\put(17.5, 0){$0$} \put(76, 0){$\tilde m$} \put(138, 0){$1$}
\end{picture}
}
\fbox{\begin{picture}(150, 50)  
\put(20, 40){\fbox{$-A^{(4)}<R<0$}}
\multiput(20, 15)(2, 0){30}{\line(1, 0){1}} \put(80, 15){\line(1,0){60}}
\put(20, 15){\circle*{5}} \put(80, 15){\circle*{5}} \put(140, 11){\line(0, 1){8}}
\put(46.5, 12.75){$<$} \put(106.5, 12.75){$<$}
\put(16, 0){$\tilde m$} \put(77.5, 0){$0$} \put(138, 0){$1$}
\end{picture}
}
\caption{Flux diagram - part II.2. The dots represent stationary points and the dashed lines indicate regions outside $[0,1]$.}
\label{partII2}
\end{figure}
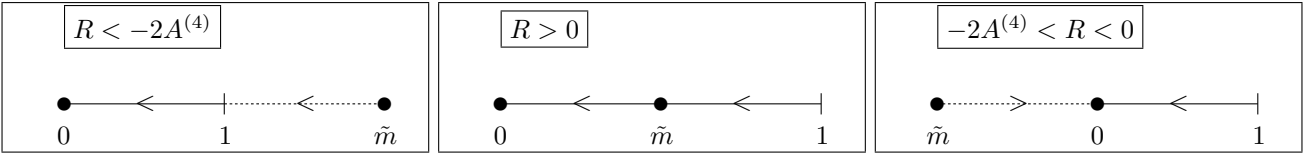

\bigskip
The differential equation for this case is
\begin{align}
\dot m = -\left(R+A^{(4)}\right)m^{2}\left(m-\tilde m\right),
\label{II2_1}
\end{align}
where $\tilde m:=\frac{R}{R+A^{(4)}}$. The system displays two different stationary points at $m^{\ast}=0$ and $m^{\ast}=\tilde m$. Since
\begin{align}
\tilde m :=\frac{R}{R+A^{(4)}} \left\{
\begin{array}{llcl}
>1        & \text{ (meaning $0<1<\tilde m$) } &,& R<-A^{(4)} \\
\in (0,1) & \text{ (meaning $0<\tilde m<1$) } &,& R>0 \\
<0        & \text{ (meaning $\tilde m<0<1$) } &,& -A^{(4)}<R<0
\end{array}
\right.,
\label{II2_2}
\end{align}
the flux diagram is depicted in figure \ref{partII2}.

This case assumes $R\neq 0$ and $R\neq -A^{(4)}$, but these two scenarios were analyzed previously (parts I.1 and I.3, respectively).

\bigskip
\noindent
\underline{II.3) Curve $R^{2}-4LA^{(4)}=0$}

\bigskip
The differential equation for this case is
\begin{align}
\dot m = -\frac{\left(R+2A^{(4)}\right)^{2}}{4A^{(4)}}m\left(m-\tilde m\right)^{2},
\label{II3_1}
\end{align}
where $\tilde m:=\frac{R}{R+2A^{(4)}}$. The system displays two different stationary points at $m^{\ast}=0$ and $m^{\ast}=\tilde m$. Since
\begin{align}
\tilde m :=\frac{R}{R+2A^{(4)}} \left\{
\begin{array}{llcl}
>1        & \text{ (meaning $0<1<\tilde m$) } &,& R<-2A^{(4)} \\
\in (0,1) & \text{ (meaning $0<\tilde m<1$) } &,& R>0 \\
<0        & \text{ (meaning $\tilde m<0<1$) } &,& -2A^{(4)}<R<0
\end{array}
\right.,
\label{II3_2}
\end{align}
the flux diagram is shown in \ref{partII3}.

\begin{figure}[ht]
\centering
\fbox{\begin{picture}(150, 50)  
\put(20, 40){\fbox{$R<-2A^{(4)}$}}
\put(20, 15){\line(1,0){60}} \multiput(80, 15)(2, 0){30}{\line(1, 0){1}}
\put(20, 15){\circle*{5}} \put(80, 11){\line(0, 1){8}} \put(140, 15){\circle*{5}}
\put(46.5, 12.75){$<$} \put(106.5, 12.75){$<$}
\put(17.5, 0){$0$} \put(78, 0){$1$} \put(136, 0){$\tilde m$}
\end{picture}
}
\fbox{\begin{picture}(150, 50)  
\put(20, 40){\fbox{$R>0$}}
\put(20, 15){\line(1,0){120}}
\put(20, 15){\circle*{5}} \put(80, 15){\circle*{5}} \put(140, 11){\line(0, 1){8}}
\put(46.5, 12.75){$<$} \put(106.5, 12.75){$<$}
\put(17.5, 0){$0$} \put(76, 0){$\tilde m$} \put(138, 0){$1$}
\end{picture}
}
\fbox{\begin{picture}(150, 50)  
\put(20, 40){\fbox{$-2A^{(4)}<R<0$}}
\multiput(20, 15)(2, 0){30}{\line(1, 0){1}} \put(80, 15){\line(1,0){60}}
\put(20, 15){\circle*{5}} \put(80, 15){\circle*{5}} \put(140, 11){\line(0, 1){8}}
\put(46.5, 12.75){$>$} \put(106.5, 12.75){$<$}
\put(16, 0){$\tilde m$} \put(77.5, 0){$0$} \put(138, 0){$1$}
\end{picture}
}
\caption{Flux diagram - part II.2. The dots represent stationary points and the dashed lines indicate regions outside $[0,1]$.}
\label{partII3}
\end{figure}

This case assumes $R\neq 0$ and $R\neq -2A^{(4)}$ and ; nevertheless, these two situations were analyzed previously (parts I.1 and I.2, respectively).


\subsection{Part III (planar regions)} 

\begin{figure}[ht!]
\centering
\includegraphics[width = 200pt]{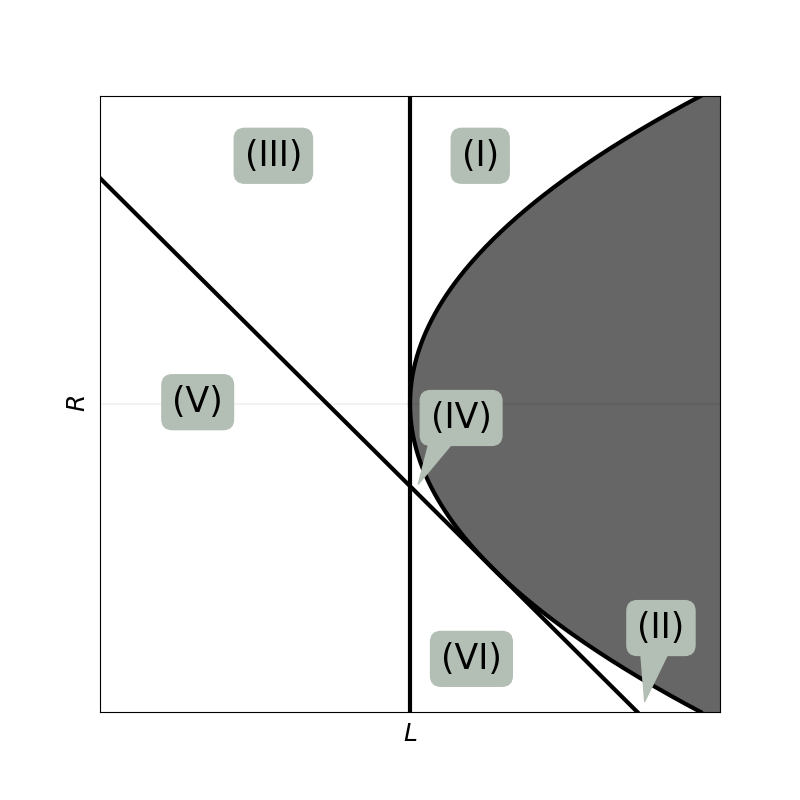}
\caption{Disjoint regions where the stability analysis is performed separately.}
\label{regions0}
\end{figure}

The regions not investigated in Part I and Part II constitute disjoint planar regions, which are displayed in figure \ref{regions0}. In the dark region (region (0) below), there is just a single stationary point, $m^{\ast}=0$, whereas the other six have two more, given by \eqref{m+-0}. In what follows, the relative positions of the points $\{0,m_{-}^{(0)},m_{+}^{(0)},1\}$ will be examined. These analysis are beneficted by the alternative representation of \eqref{m+-0}, which is
\begin{align}
m_{\pm}^{(0)} = \frac{2L}{R+2L\mp\sqrt{R^{2}-4LA^{(4)}}}.
\label{m+-0alternative}
\end{align}

\bigskip
\noindent
\underline{III.0) Region (0): $R^{2}-4LA^{(4)}<0$}

\bigskip
This is the dark area in figure \ref{regions0}, and corresponds to the situation where $m_{\pm}^{(0)}$, given in \eqref{m+-0}, are not real solutions. The differential equation for this case is
\begin{align}
\dot m = -m\left[ L - \left(R+2L\right)m + \left(R+L+A^{(4)}\right)m^{2}\right],
\label{III0_1}
\end{align}
where the expression inside the square brackets in \eqref{III0_1} is strictly positive. As stated previously, and clear from \eqref{III0_1}, there is a single stationary point, which is $m^{\ast}=0$, and the flux diagram is shown in figure \ref{partIII0}.

\begin{figure}[ht]
\centering
\fbox{\begin{picture}(150, 25)  
\put(20, 15){\line(1,0){120}}
\put(20, 15){\circle*{5}} \put(140, 11){\line(0, 1){8}}
\put(76.5, 12.75){$<$}
\put(17.5, 0){$0$} \put(138, 0){$1$}
\end{picture}
}
\caption{Flux diagram in region (0).}
\label{partIII0}
\end{figure}

\bigskip
\noindent
\underline{III.1) Region (I): $R^{2}-4LA^{(4)}>0$, $R+2L > \sqrt{R^{2}-4LA^{(4)}}$, and $R>0$}

\bigskip
From $0<\sqrt{R^{2}-4LA^{(4)}}<R+2L$, one has $L\left(R+L+A^{(4)}\right)>0$. This implies $L>0$ and $R+L+A^{(4)}>0$, since it is not possible to admit $L<0$ and $R+L+A^{(4)}<0$ at the same time: note that $R+L+A^{(4)}<0$ is equivalent to $(R+2L)+A^{(4)}<L$; the left hand-side is positive ($R+2L>\sqrt{R^{2}-4LA^{(4)}}>0$), while the right-hand side is negative (leading to an impossible situation).

From $R+2L > \sqrt{R^{2}-4LA^{(4)}}$ and $R+L+A^{(4)}>0$, it is straightforward that $0<m_{-}^{(0)}<m_{+}^{(0)}$ from \eqref{m+-0}. Furthermore, one sees that
\begin{align}
m_{+}^{(0)} = \frac{2L}{2L+\left(R-\sqrt{R^{2}-4LA^{(4)}}\right)} < 1.
\label{partIII1_1}
\end{align}
Hence, $0<m_{-}^{(0)}<m_{+}^{(0)}<1$ in region (I).

\bigskip
\noindent
\underline{III.2) Region (II): $R^{2}-4LA^{(4)}>0$, $R+2L > \sqrt{R^{2}-4LA^{(4)}}$, and $R<0$}

From the same reasoning above, one has $R+L+A^{(4)}>0$ and $L>0$, which implies $m_{-}^{(0)}<m_{+}^{(0)}$ by \eqref{m+-0}. Furthermore, from
\begin{align}
m_{-}^{(0)} = \frac{R+2L-\sqrt{R^{2}-4LA^{(4)}}}{2\left(R+L+A^{(4)}\right)} = \frac{2L}{2L-\left(-R-\sqrt{R^{2}-4LA^{(4)}}\right)},
\label{partIII2_1}
\end{align}
it follows that $m_{-}^{(0)}>0$ (from the second member of \eqref{partIII2_1}); therefore, it is straightforward that $m_{-}^{(0)}>1$ from the last member of \eqref{partIII2_1} (note that $-R-\sqrt{R^{2}-4LA^{(4)}}>0$ but the denominator of the last member is positive because $m_{-}^{(0)}>0$). Hence, $0<1<m_{-}^{(0)}<m_{+}^{(0)}$ in region (II).

\bigskip
\noindent
\underline{III.3) Region (III): $R^{2}-4LA^{(4)}>0$, $R+2L < \sqrt{R^{2}-4LA^{(4)}}$, $R+L+A^{(0)}>0$, and $L<0$}

\bigskip
From
\begin{align}
m_{+}^{(0)} = \frac{(-2L)}{(-2L)+\left(\sqrt{R^{2}-4LA^{(4)}}-R\right)},
\label{partIII3_1}
\end{align}
since $\sqrt{R^{2}-4LA^{(4)}}-R>0$ for $L<0$, then $0<m_{+}^{(0)}<1$. Furthermore,
\begin{align}
m_{-}^{(0)} = \frac{R+2L-\sqrt{R^{2}-4LA^{(4)}}}{2\left(R+L+A^{(4)}\right)} < 0,
\label{partIII3_2}
\end{align}
since the numerator and denominator do not share the same sign. Hence, $m_{-}^{(0)}<0<m_{+}^{(0)}<1$ in region (III).

\bigskip
\noindent
\underline{III.4) Region (IV): $R^{2}-4LA^{(4)}>0$, $R+2L < \sqrt{R^{2}-4LA^{(4)}}$, $R+L+A^{(0)}>0$, and $L>0$} 

\bigskip
From $R+L+A^{(0)}>0$, one has $m_{-}^{(0)}<m_{+}^{(0)}$ by \eqref{m+-0}. Furthermore,
\begin{align}
m_{+}^{(0)} = \frac{2L}{R+2L-\sqrt{R^{2}-4LA^{(4)}}} < 0,
\label{partIII4_1}
\end{align}
since the numerator and denominator do not share the same sign. Hence, $m_{-}^{(0)}<m_{+}^{(0)}<0<1$ in region (IV).

\bigskip
\noindent
\underline{III.5) Region (V): $R^{2}-4LA^{(4)}>0$, $R+2L < \sqrt{R^{2}-4LA^{(4)}}$, $R+L+A^{(0)}<0$, and $L<0$}

\bigskip
From $R+L+A^{(0)}<0$, one has $m_{+}^{(0)}<m_{-}^{(0)}$ by \eqref{m+-0}. Furthermore,
\begin{align}
m_{+}^{(0)} = \frac{(-2L)}{(-2L)+\left(\sqrt{R^{2}-4LA^{(4)}}-R\right)} \in (0,1),
\label{partIII5_1}
\end{align}
since $\sqrt{R^{2}-4LA^{(4)}}-R>0$ for $L<0$. This leads, evidently, to $m_{-}^{(0)}>0$. Finally, from
\begin{align}
m_{-}^{(0)} = \frac{(-2L)}{(-2L)-\left(\sqrt{R^{2}-4LA^{(4)}}+R\right)},
\label{partIII5_2}
\end{align}
one sees that the denominator is positive, because of $m_{-}^{(0)}>0$ and the positivity of the numerator. Since $\sqrt{R^{2}-4LA^{(4)}}+R>0$ for $L<0$, it implies $(-2L)>(-2L)-\left(\sqrt{R^{2}-4LA^{(4)}}+R\right)>0$, which leads to $m_{-}^{(0)}>1$. Hence, $0<m_{+}^{(0)}<1<m_{-}^{(0)}$ in region (V).

\bigskip
\noindent
\underline{III.6) Region (VI): $R^{2}-4LA^{(4)}>0$, $R+2L < \sqrt{R^{2}-4LA^{(4)}}$, $R+L+A^{(0)}<0$, and $L>0$} 

\bigskip
From $R+L+A^{(0)}<0$, one has $m_{+}^{(0)}<m_{-}^{(0)}$ by \eqref{m+-0}. Furthermore,
\begin{align}
m_{+}^{(0)} = \frac{2L}{R+2L-\sqrt{R^{2}-4LA^{(4)}}} < 0,
\label{partIII6_1}
\end{align}
since the numerator and denominator do not share the same sign. Moreover, using the representation \eqref{m+-0alternative}, one sees that $m_{-}^{(0)}>0$ because both the numerator and denominator have the same sign. Finally, from
\begin{align}
m_{-}^{(0)} = \frac{2L}{2L-\left(-R-\sqrt{R^{2}-4LA^{(4)}}\right)},
\label{partIII6_2}
\end{align}
one sees that the denominator is positive, because of $m_{-}^{(0)}>0$ and the positivity of the numerator. The conditions $R+L+A^{(0)}<0$ and $L>0$ implies $R<0$. Therefore, $-R-\sqrt{R^{2}-4LA^{(4)}}>0$ for $R<0$ and $L>0$, which leads to $2L>2L-\left(-R-\sqrt{R^{2}-4LA^{(4)}}\right)>0$, and, consequently, to $m_{-}^{(0)}>1$. Hence, $m_{+}^{(0)}<0<1<m_{-}^{(0)}$ in region (VI).

\bigskip

The flux diagram of Part III (except for region (0), which was already shown) is displayed in figure \ref{diagram_partIII}.

\begin{figure}[ht!]
\centering
\fbox{\begin{picture}(150, 50)  
\put(20, 40){\fbox{Region (I)}}
\put(20, 15){\line(1,0){40}} \put(60, 15){\line(1,0){40}} \put(100, 15){\line(1,0){40}} 
\put(20, 15){\circle*{5}} \put(60, 15){\circle*{5}} \put(100, 15){\circle*{5}} \put(140, 11){\line(0, 1){8}}
\put(36.5, 12.75){$<$} \put(76.5, 12.75){$>$} \put(116.5, 12.75){$<$}
\put(17.5, 0){$0$} \put(53, 1){\scriptsize$m_{-}^{(0)}$} \put(93, 1){\scriptsize$m_{+}^{(0)}$} \put(138, 0){$1$}
\end{picture}
}
\fbox{\begin{picture}(150, 50)  
\put(20, 40){\fbox{Region (II)}}
\put(20, 15){\line(1,0){40}} \multiput(60, 15)(2, 0){20}{\line(1, 0){1}} \multiput(100, 15)(2, 0){20}{\line(1, 0){1}}
\put(20, 15){\circle*{5}} \put(60, 11){\line(0, 1){8}} \put(100, 15){\circle*{5}} \put(140, 15){\circle*{5}}
\put(17.5, 0){$0$} \put(58, 0){$1$} \put(93, 1){\scriptsize$m_{-}^{(0)}$} \put(133, 1){\scriptsize$m_{+}^{(0)}$} 
\put(36.5, 12.75){$<$} \put(76.5, 12.75){$<$} \put(116.5, 12.75){$>$}
\end{picture}
}
\fbox{\begin{picture}(150, 50)  
\put(20, 40){\fbox{Region (III)}}
\multiput(20, 15)(2, 0){20}{\line(1, 0){1}} \put(60, 15){\line(1,0){40}} \put(100, 15){\line(1,0){40}} 
\put(20, 15){\circle*{5}} \put(60, 15){\circle*{5}} \put(100, 15){\circle*{5}} \put(140, 11){\line(0, 1){8}}
\put(13, 1){\scriptsize$m_{-}^{(0)}$} \put(57.5, 0){$0$} \put(93, 1){\scriptsize$m_{+}^{(0)}$} \put(138, 0){$1$}
\put(36.5, 12.75){$<$} \put(76.5, 12.75){$>$} \put(116.5, 12.75){$<$}
\end{picture}
}
\\
\fbox{\begin{picture}(150, 50)  
\put(20, 40){\fbox{Region (IV)}}
\multiput(20, 15)(2, 0){20}{\line(1, 0){1}} \multiput(60, 15)(2, 0){20}{\line(1, 0){1}} \put(100, 15){\line(1,0){40}} 
\put(20, 15){\circle*{5}} \put(60, 15){\circle*{5}} \put(100, 15){\circle*{5}} \put(140, 11){\line(0, 1){8}}
\put(13, 1){\scriptsize$m_{-}^{(0)}$} \put(53, 1){\scriptsize$m_{+}^{(0)}$} \put(97.5, 0){$0$} \put(138, 0){$1$}
\put(36.5, 12.75){$<$} \put(76.5, 12.75){$>$} \put(116.5, 12.75){$<$}
\end{picture}
}
\fbox{\begin{picture}(150, 50)  
\put(20, 40){\fbox{Region (V)}}
\put(20, 15){\line(1,0){40}} \put(60, 15){\line(1,0){40}} \multiput(100, 15)(2, 0){20}{\line(1, 0){1}}
\put(20, 15){\circle*{5}} \put(60, 15){\circle*{5}} \put(100, 11){\line(0, 1){8}} \put(140, 15){\circle*{5}}
\put(17.5, 0){$0$} \put(53, 1){\scriptsize$m_{+}^{(0)}$} \put(98, 0){$1$} \put(133, 1){\scriptsize$m_{-}^{(0)}$} 
\put(36.5, 12.75){$>$} \put(76.5, 12.75){$<$} \put(116.5, 12.75){$<$}
\end{picture}
}
\fbox{\begin{picture}(150, 50)  
\put(20, 40){\fbox{Region (VI)}}
\multiput(20, 15)(2, 0){20}{\line(1, 0){1}} \put(60, 15){\line(1,0){40}} \multiput(100, 15)(2, 0){20}{\line(1, 0){1}}
\put(20, 15){\circle*{5}} \put(60, 15){\circle*{5}} \put(100, 11){\line(0, 1){8}} \put(140, 15){\circle*{5}}
\put(13, 1){\scriptsize$m_{+}^{(0)}$} \put(57.5, 0){$0$} \put(98, 0){$1$} \put(133, 1){\scriptsize$m_{-}^{(0)}$} 
\put(36.5, 12.75){$>$} \put(76.5, 12.75){$<$} \put(116.5, 12.75){$<$}
\end{picture}
}
\caption{Flux diagram - part III. The dots represent stationary points and the dashed lines indicate regions outside $[0,1]$.}
\label{diagram_partIII}
\end{figure}


\section{Appendix: case $\overline{A^{(1)}}>0$ and $A^{(4)}=0$}
\label{caseII}
\setcounter{equation}{0}

In this case, the absorbing state $m^{\ast}=1$ is predicted by the model, while the stationary point $m^{\ast}=0$ is excluded by the condition $\overline{A^{(1)}}\neq 0$. In principle, this case would follow the same reasoning as in the previous appendix, but a shortcut is available exploring the symmetry of the model. By introducing the variable
\begin{align}
n(t) := 1 - m(t),
\label{n}
\end{align}
which also belongs to the interval $[0,1]$, the dynamical system $\dot m=F_{1}(m)$, where $F_{1}(m)$ is given in \eqref{F1}, is cast as
\begin{align}
\dot n = -n\left[ R - \left(2R+L\right)n + \left(L+R+\overline{A^{(1)}}\right)n^{2} \right],
\label{dndt}
\end{align}
where the right-hand side is similar to $F_{0}(m)$ by exchanging $R\leftrightarrow L$ and $A^{(4)}\leftrightarrow \overline{A^{(1)}}$. Furthermore, the roots of the polynomial inside the square brackets in \eqref{dndt} are
\begin{align}
n_{\pm} := \frac{2R + L \pm\sqrt{ L^{2} - 4R\overline{A^{(1)}} } }{2\left(R + L + \overline{A^{(1)}}\right)},
\label{n+-}
\end{align}
which is also similar to $m_{\pm}^{(0)}$ in \eqref{m+-0} by exchanging $R\leftrightarrow L$ and $A^{(4)}\leftrightarrow \overline{A^{(1)}}$. It is also important to notice that these two sets of stationary solutions are related by
\begin{align}
m_{\mp}^{(1)} + n_{\pm} = 1.
\label{nm1}
\end{align}

Hereby, one may benefit from the previous appendix by adapting the results obtained there to the present case, and then express the outcomes in terms of the variables specific to this case. The analysis is again separated into disjoint regions specified in table \ref{parts1}, and the results are listed for reference.

\begin{table}[ht!]
\centering
\renewcommand*{\arraystretch}{1.4}
\begin{tabular}{|c|l|}\hline
\multirow{3}{*}{Part IV} & Point $(0,0)$ \\\cline{2-2}
 & Point $(-2\overline{A^{(1)}}, \overline{A^{(1)}})$ \\\cline{2-2}
 & Point $(-\overline{A^{(1)}},0)$ \\\hline
\multirow{3}{*}{Part V} & Line $R+L+\overline{A^{(1)}}=0$ \\\cline{2-2}
 & Line $R=0$ \\\cline{2-2}
 & Curve $L^{2}-4R\overline{A^{(1)}}=0$ \\\hline
\multirow{6}{*}{Part VI} & Region (0) \\\cline{2-2}
 & Region (I) \\\cline{2-2}
 & Region (II) \\\cline{2-2}
 & Region (III) \\\cline{2-2}
 & Region (IV) \\\cline{2-2}
 & Region (V) \\\cline{2-2}
 & Region (VI) \\\hline
\end{tabular}
\caption{Division of the $(L, R)$ plane suitable for the construction of the flux diagram of the case $\overline{A^{(1)}}>0$ and $A^{(4)}=0$; the regions in Part VI are defined in Fig. \ref{regions1}.}
\label{parts1}
\end{table}


\subsection{Part IV (points)}

The flux diagram of part IV is given in table \ref{diagram_partIV}.

\begin{table}[ht!]
\centering
\renewcommand*{\arraystretch}{1.4}
\begin{tabular}{|c|l|c|}\hline
Point                                     & $F_{1}(m)$                              & Flux diagram \\\hline
$(0, 0)$                                  & $\overline{A^{(1)}}\left(1-m\right)^{3}$ & \multirow{3}{*}{ \begin{picture}(120,30)\put(20, 15){\line(1,0){80}} \put(20, 11){\line(0, 1){8}} \put(100, 15){\circle*{5}} \put(17.5, 0){$0$} \put(98, 0){$1$} \put(56.5, 12.75){$>$} \end{picture} } \\\cline{1-2}
$(-2\overline{A^{(1)}},\overline{A^{(1)}})$ & $\overline{A^{(1)}}\left(1-m\right)$ & \\\cline{1-2}
$(-\overline{A^{(1)}},0)$                  & $\overline{A^{(1)}}\left(1-m\right)^{2}$ & \\\hline
\end{tabular}
\caption{Flux diagram for the points in part IV.}
\label{diagram_partIV}
\end{table}


\subsection{Part V (curves)}

The flux diagrams for the lines and curves in part V are given in table \ref{diagram_tableV}

\begin{table}[ht!]
\centering
\renewcommand*{\arraystretch}{1.4}
\begin{tabular}{|c|c|c|c|c|}\hline
Curve & $F_{1}(m)$ & $\tilde m^{(1)}$ & Condition & Flux diagram \\\hline
\multirow{3}{*}{ \parbox{70pt}{$R+L+\overline{A^{(1)}}=0$} } & \multirow{3}{*}{\parbox{140pt}{$\big(\overline{A^{(1)}}-R\big)\big(1-m\big)\big(\tilde m^{(1)}-m\big)$}} & \multirow{3}{*}{\parbox{40pt}{$\frac{\overline{A^{(1)}}}{\overline{A^{(1)}}-R}$}} & \parbox{60pt}{$R<0$} & \parbox{120pt}{ \begin{picture}(120, 30) \put(20, 20){\line(1,0){80}} \put(20, 16){\line(0, 1){8}} \put(60, 20){\circle*{5}} \put(100, 20){\circle*{5}} \put(17.5, 5){$0$} \put(53, 5){\scriptsize$\tilde m^{(1)}$} \put(98, 5){$1$} \put(36.5, 17.75){$>$} \put(76.5, 17.75){$<$} \end{picture}} \\\cline{4-5}
 & & & \multirow{1}{*}{\parbox{60pt}{$0<R<\overline{A^{(1)}}$}} & \parbox{120pt}{ \begin{picture}(120,30)\put(20, 20){\line(1,0){40}} \multiput(60, 20)(2, 0){20}{\line(1, 0){1}} \put(20, 16){\line(0, 1){8}} \put(60, 20){\circle*{5}} \put(100, 20){\circle*{5}} \put(17.5, 5){$0$} \put(98, 5){$1$} \put(53, 5){\scriptsize$\tilde m^{(1)}$} \put(36.5, 17.75){$>$} \put(76.5, 17.75){$<$} \end{picture}} \\\cline{4-5}
 & & & \multirow{1}{*}{\parbox{70pt}{$R>\overline{A^{(1)}}$}} & \parbox{120pt}{ \begin{picture}(120,30) \multiput(20, 20)(2, 0){20}{\line(1, 0){1}} \put(60, 20){\line(1,0){40}} \put(20, 20){\circle*{5}} \put(60, 16){\line(0, 1){8}}  \put(100, 20){\circle*{5}} \put(13, 5){\scriptsize$\tilde m^{(1)}$} \put(57.5, 5){$0$} \put(98, 5){$1$} \put(36.5, 17.75){$>$} \put(76.5, 17.75){$>$} \end{picture}} \\\hline
\multirow{3}{*}{ \parbox{70pt}{$R=0$} } & \multirow{3}{*}{\parbox{140pt}{$\big(L+\overline{A^{(1)}}\big)\big(1-m\big)^{2}\big(\tilde m^{(1)}-m\big)$}} & \multirow{3}{*}{\parbox{40pt}{$\frac{\overline{A^{(1)}}}{\overline{L+A^{(1)}}}$}} & \parbox{60pt}{$L<-\overline{A^{(1)}}$} & \parbox{120pt}{ \begin{picture}(120, 30) \multiput(20, 20)(2, 0){20}{\line(1, 0){1}} \put(60, 20){\line(1,0){40}} \put(20, 20){\circle*{5}} \put(60, 16){\line(0, 1){8}} \put(100, 20){\circle*{5}} \put(13, 5){\scriptsize$\tilde m^{(1)}$} \put(57.5, 5){$0$} \put(98, 5){$1$} \put(36.5, 17.75){$>$} \put(76.5, 17.75){$>$} \end{picture}} \\\cline{4-5}
 & & & \multirow{1}{*}{\parbox{60pt}{$-\overline{A^{(1)}}<L<0$}} & \parbox{120pt}{ \scalebox{1}{\begin{picture}(120,30) \put(20, 20){\line(1,0){40}} \multiput(60, 20)(2, 0){20}{\line(1, 0){1}} \put(20, 16){\line(0, 1){8}} \put(60, 20){\circle*{5}} \put(100, 20){\circle*{5}} \put(17.5, 5){$0$} \put(58, 5){$1$} \put(93, 5){\scriptsize$\tilde m^{(1)}$} \put(36.5, 17.75){$>$} \put(76.5, 17.75){$>$} \end{picture}}} \\\cline{4-5}
 & & & \multirow{1}{*}{\parbox{70pt}{$L>0$}} & \parbox{120pt}{ \scalebox{1}{\begin{picture}(120,30) \put(20, 20){\line(1,0){80}} \put(20, 16){\line(0, 1){8}} \put(60, 20){\circle*{5}} \put(100, 20){\circle*{5}} \put(17.5, 5){$0$} \put(53, 5){\scriptsize$\tilde m^{(1)}$} \put(98, 5){$1$} \put(36.5, 17.75){$>$} \put(76.5, 17.75){$<$} \end{picture}}} \\\hline
\multirow{3}{*}{ \parbox{70pt}{$L^{2}-4R\overline{A^{(1)}}=0$} } & \multirow{3}{*}{\parbox{140pt}{$\frac{\big(2\overline{A^{(1)}}+L\big)^{2}}{4\overline{A^{(1)}}}\big(1-m\big)\big(\tilde m^{(1)}-m\big)^{2}$}} & \multirow{3}{*}{\parbox{40pt}{$\frac{\overline{2A^{(1)}}}{\overline{2A^{(1)}}+L}$}} & \parbox{60pt}{$L<-2\overline{A^{(1)}}$} & \parbox{120pt}{ \begin{picture}(120, 30) \multiput(20, 20)(2, 0){20}{\line(1, 0){1}} \put(60, 20){\line(1,0){40}} \put(20, 20){\circle*{5}} \put(60, 16){\line(0, 1){8}} \put(100, 20){\circle*{5}} \put(13, 5){\scriptsize$\tilde m^{(1)}$} \put(57.5, 5){$0$} \put(98, 5){$1$} \put(36.5, 17.75){$>$} \put(76.5, 17.75){$>$} \end{picture}} \\\cline{4-5}
 & & & \multirow{1}{*}{\parbox{70pt}{$-2\overline{A^{(1)}}<L<0$}} & \parbox{120pt}{ \begin{picture}(120,30) \put(20, 20){\line(1,0){40}} \multiput(60, 20)(2, 0){20}{\line(1, 0){1}} \put(20, 16){\line(0, 1){8}} \put(60, 20){\circle*{5}} \put(100, 20){\circle*{5}} \put(17.5, 5){$0$} \put(58, 5){$1$} \put(93, 5){\scriptsize$\tilde m^{(1)}$} \put(36.5, 17.75){$>$} \put(76.5, 17.75){$<$} \end{picture}} \\\cline{4-5}
 & & & \multirow{1}{*}{\parbox{60pt}{$L>0$}} & \parbox{120pt}{ \begin{picture}(120,30) \put(20, 20){\line(1,0){80}} \put(20, 16){\line(0, 1){8}} \put(60, 20){\circle*{5}} \put(100, 20){\circle*{5}} \put(17.5, 5){$0$} \put(53, 5){\scriptsize$\tilde m^{(1)}$} \put(98, 5){$1$} \put(36.5, 17.75){$>$} \put(76.5, 17.75){$>$} \end{picture}} \\\hline
\end{tabular}
\caption{Flux diagram for the lines and curve of part V.}
\label{diagram_tableV}
\end{table}


\subsection{Part VI (planar regions)}

\begin{figure}[ht!]
\centering
\includegraphics[width = 200pt]{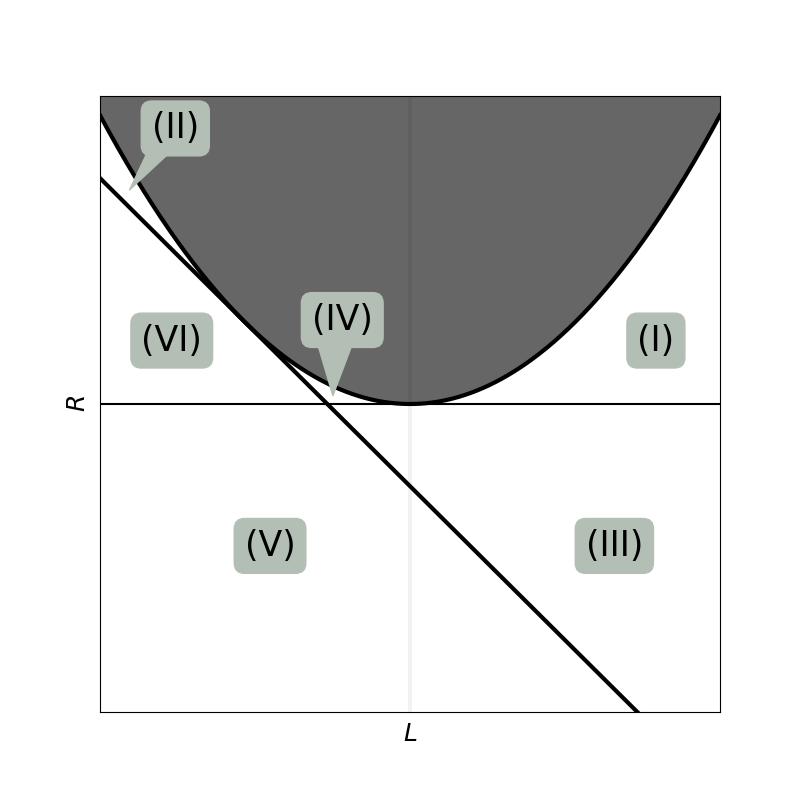}
\caption{Disjoint regions where the stability analysis is performed separately.}
\label{regions1}
\end{figure}

The regions indicated in part VI are shown in Fig. \ref{regions1}.


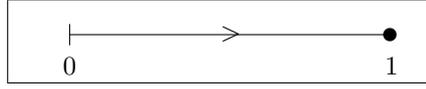
\begin{figure}[ht]
\centering
\fbox{\begin{picture}(150, 25)  
\put(20, 15){\line(1,0){120}}
\put(20, 11){\line(0, 1){8}} \put(140, 15){\circle*{5}}
\put(76.5, 12.75){$>$}
\put(17.5, 0){$0$} \put(138, 0){$1$}
\end{picture}
}
\caption{Flux diagram in region (0).}
\label{partVI0}
\end{figure}

\begin{figure}[ht!]
\centering
\fbox{\begin{picture}(150, 50)  
\put(20, 40){\fbox{Region (I)}}
\put(20, 15){\line(1,0){40}} \put(60, 15){\line(1,0){40}} \put(100, 15){\line(1,0){40}} 
\put(20, 11){\line(0, 1){8}} \put(60, 15){\circle*{5}} \put(100, 15){\circle*{5}} \put(140, 15){\circle*{5}}  
\put(36.5, 12.75){$>$} \put(76.5, 12.75){$<$} \put(116.5, 12.75){$>$}
\put(17.5, 0){$0$} \put(53, 1){\scriptsize$m_{-}^{(1)}$} \put(93, 1){\scriptsize$m_{+}^{(1)}$} \put(138, 0){$1$}
\end{picture}
}
\fbox{\begin{picture}(150, 50)  
\put(20, 40){\fbox{Region (II)}}
\multiput(20, 15)(2, 0){20}{\line(1, 0){1}} \multiput(60, 15)(2, 0){20}{\line(1, 0){1}} \put(100, 15){\line(1,0){40}} 
\put(20, 15){\circle*{5}} \put(60, 15){\circle*{5}} \put(100, 11){\line(0, 1){8}} \put(140, 15){\circle*{5}} 
\put(13, 1){\scriptsize$m_{-}^{(1)}$} \put(53, 1){\scriptsize$m_{+}^{(1)}$} \put(97.5, 0){$0$} \put(138, 0){$1$}
\put(36.5, 12.75){$<$} \put(76.5, 12.75){$>$} \put(116.5, 12.75){$>$}
\end{picture}
}
\fbox{\begin{picture}(150, 50)  
\put(20, 40){\fbox{Region (III)}}
\put(20, 15){\line(1,0){40}} \put(60, 15){\line(1,0){40}} \multiput(100, 15)(2, 0){20}{\line(1, 0){1}}
\put(20, 11){\line(0, 1){8}} \put(60, 15){\circle*{5}} \put(100, 15){\circle*{5}} \put(140, 15){\circle*{5}}
\put(17.5, 0){$0$} \put(53, 1){\scriptsize$m_{-}^{(1)}$} \put(98, 0){$1$} \put(133, 1){\scriptsize$m_{+}^{(1)}$} 
\put(36.5, 12.75){$>$} \put(76.5, 12.75){$<$} \put(116.5, 12.75){$>$}
\end{picture}
}
\\
\fbox{\begin{picture}(150, 50)  
\put(20, 40){\fbox{Region (IV)}}
\put(20, 15){\line(1,0){40}} \multiput(60, 15)(2, 0){20}{\line(1, 0){1}} \multiput(100, 15)(2, 0){20}{\line(1, 0){1}}
\put(20, 11){\line(0, 1){8}} \put(60, 15){\circle*{5}} \put(100, 15){\circle*{5}} \put(140, 15){\circle*{5}}
\put(17.5, 0){$0$} \put(58, 0){$1$} \put(93, 1){\scriptsize$m_{-}^{(1)}$} \put(133, 1){\scriptsize$m_{+}^{(1)}$} 
\put(36.5, 12.75){$>$} \put(76.5, 12.75){$<$} \put(116.5, 12.75){$>$}
\end{picture}
}
\fbox{\begin{picture}(150, 50)  
\put(20, 40){\fbox{Region (V)}}
\multiput(20, 15)(2, 0){20}{\line(1, 0){1}} \put(60, 15){\line(1,0){40}} \put(100, 15){\line(1,0){40}} 
\put(20, 15){\circle*{5}} \put(60, 11){\line(0, 1){8}} \put(100, 15){\circle*{5}} \put(140, 15){\circle*{5}}
\put(13, 1){\scriptsize$m_{+}^{(1)}$} \put(57.5, 0){$0$} \put(93, 1){\scriptsize$m_{-}^{(1)}$} \put(138, 0){$1$}
\put(36.5, 12.75){$>$} \put(76.5, 12.75){$>$} \put(116.5, 12.75){$<$}
\end{picture}
}
\fbox{\begin{picture}(150, 50)  
\put(20, 40){\fbox{Region (VI)}}
\multiput(20, 15)(2, 0){20}{\line(1, 0){1}} \put(60, 15){\line(1,0){40}} \multiput(100, 15)(2, 0){20}{\line(1, 0){1}}
\put(20, 15){\circle*{5}} \put(60, 11){\line(0, 1){8}} \put(100, 15){\circle*{5}} \put(140, 15){\circle*{5}}
\put(13, 1){\scriptsize$m_{+}^{(1)}$} \put(57.5, 0){$0$} \put(98, 0){$1$} \put(133, 1){\scriptsize$m_{-}^{(1)}$} 
\put(36.5, 12.75){$>$} \put(76.5, 12.75){$>$} \put(116.5, 12.75){$<$}
\end{picture}
}
\caption{Flux diagrams for the regions in part VI.}
\label{diagram_partVI}
\end{figure}
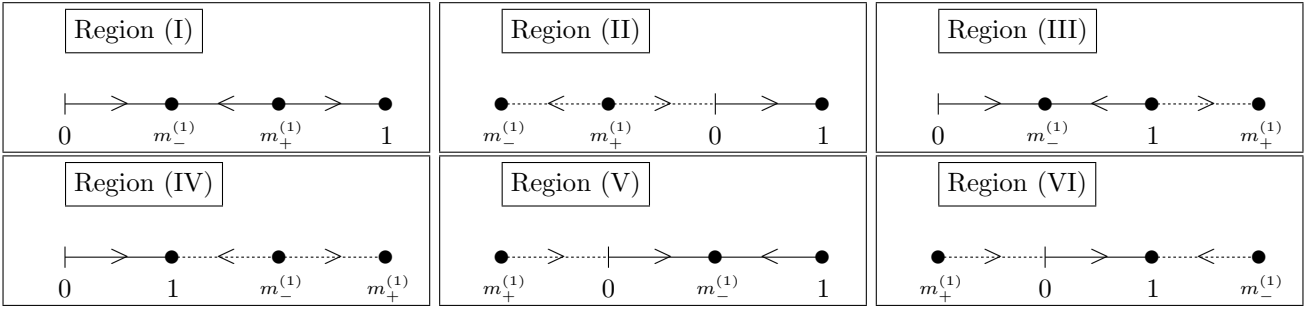

The dark area in figure \ref{regions1}, and corresponds to the situation where $m_{\pm}^{(1)}$, given in \eqref{m+-1}, are not real solutions. The differential equation for this case is
\begin{align}
\dot m = \left(1-m\right)\left[ \overline{A^{(1)}} - \left(2\overline{A^{(1)}}+L\right)m + \left(R+L+\overline{A^{(1)}}\right)m^{2} \right],
\label{VI1_1}
\end{align}
where the expression inside the square brackets in \eqref{VI1_1} is strictly positive. As stated previously, and clear from \eqref{VI1_1}, there is a single stationary point, which is $m^{\ast}=1$, and the flux diagram is shown in figure \ref{partVI0}. The flux diagram of the regions of part VI is shown in Fig. \ref{diagram_partVI}.

The remaining regions are analyzed below.

\section{Appendix: case $\overline{A^{(1)}}=0$ and $A^{(4)}=0$}
\label{caseIII}
\setcounter{equation}{0}

In this last scenario, where $\overline{A^{(1)}}=0$ and $A^{(4)}=0$, the equation of motion can be cast as $\dot m = F_{01}(m)$, where
\begin{align}
\nonumber F_{01}(m) &:= -Lm + \left(R+2L\right)m^{2} - \left(R+L\right)m^{3} \\
 &= m\left(1-m\right)\left[ \left(R+L\right)m - L \right].
\label{F01appendix}
\end{align}
Apart from the stationary points $m^{\ast}=0$ and $m^{\ast}=1$, a third one, $\tilde m^{(01)}$, emerges depending on the region $(L,R)$, and is given by
\begin{align}
\tilde m^{(01)} := \frac{L}{R+L}.
\label{m01}
\end{align}

The analysis is separated into the parts shown in Table \ref{parts1}
\begin{table}[ht]
\centering
\begin{tabular}{|c|l|}\hline
\multirow{3}{*}{Part VII} & Line $R=0$ \\\cline{2-2}
 & Line $L=0$ \\\cline{2-2}
 & Line $R+L=0$ \\\hline
\multirow{6}{*}{Part VIII} & Region (I) \\\cline{2-2}
 & Region (II) \\\cline{2-2}
 & Region (III) \\\cline{2-2}
 & Region (IV) \\\cline{2-2}
 & Region (V) \\\cline{2-2}
 & Region (VI) \\\hline
\end{tabular}
\caption{Division of the $(L, R)$ plane suitable for the construction of the flux diagram of the case $\overline{A^{(1)}}=0$ and $A^{(4)}=0$; the regions are defined in Fig. \ref{regions01}.}
\label{parts01}
\end{table}


\subsection{Part VII (lines)}

In Part VII, the stationary points are $m^{\ast}=0$ or $m^{\ast}=1$ only. The flux diagrams of all cases in this part are depicted in table \ref{diagram_tableVII}.

\begin{table}[ht!]
\centering
\renewcommand*{\arraystretch}{1.0}
\begin{tabular}{|c|c|c|c|}\hline
Line & $F_{01}(m)$ & Condition & Flux diagram \\\hline
\multirow{2}{*}{ \parbox{40pt}{$R=0$} } & \multirow{2}{*}{\parbox{80pt}{$-Lm\left(1-m\right)^{2}$}} & \parbox{60pt}{$L>0$} & \parbox{140pt}{ \begin{picture}(130, 30) \put(20, 20){\line(1,0){90}} \put(20, 20){\circle*{5}} \put(110, 20){\circle*{5}} \put(17.5, 5){$0$} \put(107.5, 5){$1$} \put(61.5, 17.75){$<$} \end{picture}} \\\cline{3-4}
 & & \multirow{1}{*}{\parbox{60pt}{$L<0$}} & \parbox{145pt}{ \begin{picture}(130,30)\put(20, 20){\line(1,0){90}} \put(20, 20){\circle*{5}} \put(110, 20){\circle*{5}} \put(17.5, 5){$0$} \put(107, 5){$1$} \put(61.5, 17.75){$>$} \end{picture}} \\\hline
\multirow{2}{*}{ \parbox{40pt}{$L=0$} } & \multirow{2}{*}{\parbox{80pt}{$Rm^{2}\left(1-m\right)$}} & \parbox{60pt}{$R>0$} & \parbox{140pt}{ \begin{picture}(130, 30) \put(20, 20){\line(1,0){90}} \put(20, 20){\circle*{5}} \put(110, 20){\circle*{5}} \put(17.5, 5){$0$} \put(107.5, 5){$1$} \put(61.5, 17.75){$>$} \end{picture}} \\\cline{3-4}
 & & \multirow{1}{*}{\parbox{60pt}{$R<0$}} & \parbox{145pt}{ \begin{picture}(130,30)\put(20, 20){\line(1,0){90}} \put(20, 20){\circle*{5}} \put(110, 20){\circle*{5}} \put(17.5, 5){$0$} \put(107, 5){$1$} \put(61.5, 17.75){$<$} \end{picture}} \\\hline
\multirow{2}{*}{ \parbox{40pt}{$R+L=0$} } & \multirow{2}{*}{\parbox{80pt}{$-Lm\left(1-m\right)$}} & \parbox{60pt}{$L>0$} & \parbox{140pt}{ \begin{picture}(130, 30) \put(20, 20){\line(1,0){90}} \put(20, 20){\circle*{5}} \put(110, 20){\circle*{5}} \put(17.5, 5){$0$} \put(107.5, 5){$1$} \put(61.5, 17.75){$<$} \end{picture}} \\\cline{3-4}
 & & \multirow{1}{*}{\parbox{60pt}{$L<0$}} & \parbox{145pt}{ \begin{picture}(130,30)\put(20, 20){\line(1,0){90}} \put(20, 20){\circle*{5}} \put(110, 20){\circle*{5}} \put(17.5, 5){$0$} \put(107, 5){$1$} \put(61.5, 17.75){$>$} \end{picture}} \\\hline
\end{tabular}
\caption{Flux diagram for the lines and curve of part V.}
\label{diagram_tableVII}
\end{table}


\subsection{Part VIII (planar regions)}

\begin{figure}[ht!]
\centering
\includegraphics[width = 200pt]{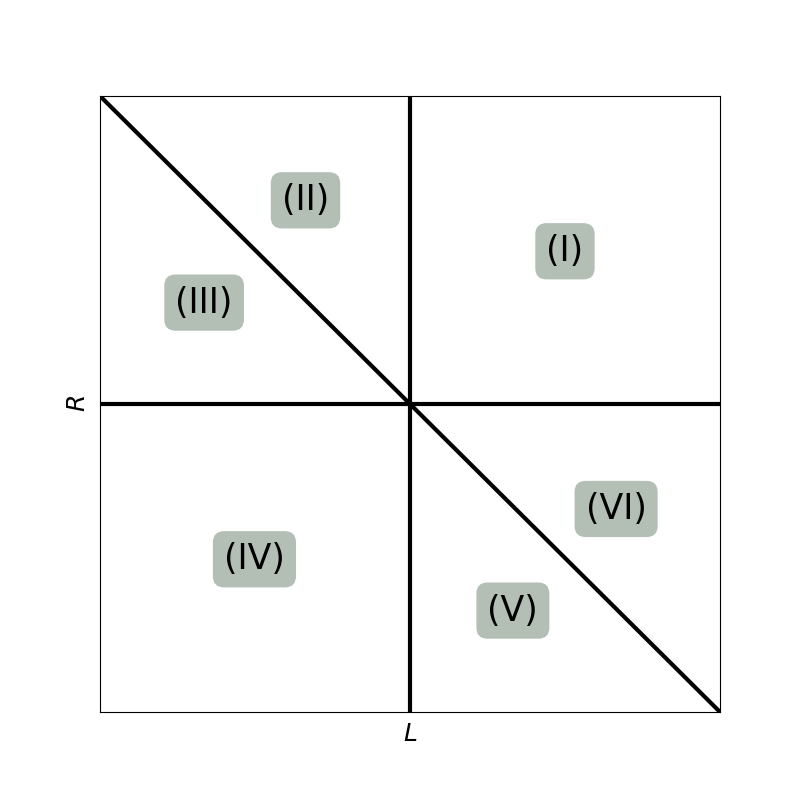}
\caption{Disjoint regions where the stability analysis is performed separately.}
\label{regions01}
\end{figure}

The stability analysis of part VIII consists of analyzing six disjoint regions indicated in Fig. \ref{regions01} (see Table \ref{parts01}).

In these regions, a third stationary point (apart from $m^{\ast}=0$ and $m^{\ast}=1$), $\tilde m^{(01)}$, given by \eqref{m01}, emerges. It is straightforward that
\begin{align}
m^{(01)} = \frac{L}{R+L} = \left\{
\begin{array}{lcl}
>1 &,& (R+L>0\textnormal{ and }R<0) \textnormal{ or }(R+L<0\textnormal{ and }R>0) \\
\in(0,1) &,& (R+L>0\textnormal{ and }R,L>0) \textnormal{ or }(R+L<0\textnormal{ and }R,L<0) \\
<0 &,& (R+L>0\textnormal{ and }L<0) \textnormal{ or }(R+L<0\textnormal{ and }L>0)
\end{array}
\right.,
\label{m01_conditions}
\end{align}
and this characterization of $\tilde m^{(01)}$ implies the flux diagrams presented in figure \ref{diagrams_partVIII}.

\begin{figure}[ht!]
\centering
\fbox{\begin{picture}(150, 50)  
\put(20, 40){\fbox{Region (I)}}
\put(20, 15){\line(1,0){60}} \put(80, 15){\line(1,0){60}}
\put(20, 15){\circle*{5}} \put(80, 15){\circle*{5}} \put(140, 15){\circle*{5}}  
\put(46.5, 12.75){$<$} \put(106.5, 12.75){$>$}
\put(17.5, 0){$0$} \put(76, 1){\scriptsize$\tilde m$} \put(138, 0){$1$}
\end{picture}
}
\fbox{\begin{picture}(150, 50)  
\put(20, 40){\fbox{Region (II)}}
\multiput(20, 15)(2, 0){30}{\line(1, 0){1}} \put(80, 15){\line(1,0){60}}
\put(20, 15){\circle*{5}} \put(80, 15){\circle*{5}} \put(140, 15){\circle*{5}}  
\put(46.5, 12.75){$<$} \put(106.5, 12.75){$>$}
\put(16, 1){\scriptsize$\tilde m$} \put(77.5, 0){$0$} \put(138, 0){$1$}
\end{picture}
}
\fbox{\begin{picture}(150, 50)  
\put(20, 40){\fbox{Region (III)}}
\put(20, 15){\line(1,0){60}} \multiput(80, 15)(2, 0){30}{\line(1, 0){1}}
\put(20, 15){\circle*{5}} \put(80, 15){\circle*{5}} \put(140, 15){\circle*{5}}  
\put(46.5, 12.75){$>$} \put(106.5, 12.75){$<$}
\put(17.5, 0){$0$} \put(78, 0){$1$} \put(136, 1){\scriptsize$\tilde m$}
\end{picture}
}
\\
\fbox{\begin{picture}(150, 50)  
\put(20, 40){\fbox{Region (IV)}}
\put(20, 15){\line(1,0){60}} \put(80, 15){\line(1,0){60}}
\put(20, 15){\circle*{5}} \put(80, 15){\circle*{5}} \put(140, 15){\circle*{5}}  
\put(46.5, 12.75){$>$} \put(106.5, 12.75){$<$}
\put(17.5, 0){$0$} \put(76, 1){\scriptsize$\tilde m$} \put(138, 0){$1$}
\end{picture}
}
\fbox{\begin{picture}(150, 50)  
\put(20, 40){\fbox{Region (V)}}
\multiput(20, 15)(2, 0){30}{\line(1, 0){1}} \put(80, 15){\line(1,0){60}}
\put(20, 15){\circle*{5}} \put(80, 15){\circle*{5}} \put(140, 15){\circle*{5}}  
\put(46.5, 12.75){$>$} \put(106.5, 12.75){$<$}
\put(16, 1){\scriptsize$\tilde m$} \put(77.5, 0){$0$} \put(138, 0){$1$}
\end{picture}
}
\fbox{\begin{picture}(150, 50)  
\put(20, 40){\fbox{Region (VI)}}
\put(20, 15){\line(1,0){60}} \multiput(80, 15)(2, 0){30}{\line(1, 0){1}}
\put(20, 15){\circle*{5}} \put(80, 15){\circle*{5}} \put(140, 15){\circle*{5}}  
\put(46.5, 12.75){$<$} \put(106.5, 12.75){$>$}
\put(17.5, 0){$0$} \put(78, 0){$1$} \put(136, 1){\scriptsize$\tilde m$}
\end{picture}
}
\caption{Flux diagram - part VIII. The dots represent stationary points and the dashed lines indicate regions outside $[0,1]$.}
\label{diagrams_partVIII}
\end{figure}



\end{document}